\documentclass[pra, preprint, superscriptaddress, nofootinbib, showpacs]{revtex4}
\usepackage{amsmath}
\usepackage{amsfonts, amssymb}
\usepackage{dsfont}
\usepackage{graphicx}
\usepackage{subfigure}
\usepackage{xcolor}

\def\comment#1{}

\newcommand{\be}{\begin{equation}}\newcommand{\ee}{\end{equation}}
\newcommand{\bea}{\begin{eqnarray}}\newcommand{\eea}{\end{eqnarray}}
\newcommand{\beaa}{\begin{eqnarray}}\newcommand{\eeaa}{\end{eqnarray}}
\newcommand{\ba}{\begin{array}}\newcommand{\ea}{\end{array}}
\newcommand{\bit}{\begin{itemize}}\newcommand{\eit}{\end{itemize}}
\newcommand{\ben}{\begin{enumerate}}\newcommand{\een}{\end{enumerate}}

\newcommand\cev[1]{\overleftarrow{#1}}

									\def\be{\begin{equation}}
\def\ee{\end{equation}}				\def\bea{\begin{eqnarray}}				\def\eea{\end{eqnarray}}
\def\bear{\begin{array}}				\def\eear{\end{array}}					
						\def\pa{\partial}							
									    
												\def\h{\eta}
													
															\def\s{\sigma}
						\def\t{\tau}								
														\def\p{\pi}
\def\a{\alpha}												
\def\d{\delta}																			
												\def\D{\Delta}
\def\w{\omega}						\def\W{\Omega}						\def\Y{\Psi}
												
\def\nn{\nonumber}											
												
												\def\x5{x^{5}}

													\def\fr{\frac}

\begin{document}

\title{Calculation of photoelectron spectra within the time-dependent configuration interaction singles scheme}

\author{Antonia Karamatskou}
\email{antonia.karamatskou@cfel.de}
\affiliation{Center for Free-Electron Laser Science, DESY, D-22607 Hamburg, Germany}
\affiliation{Department of Physics, University of Hamburg, D-20355 Hamburg, Germany}

\author{Stefan Pabst}
\affiliation{Center for Free-Electron Laser Science, DESY, D-22607 Hamburg, Germany}

\author{Yi-Jen Chen}
\affiliation{Center for Free-Electron Laser Science, DESY, D-22607 Hamburg, Germany}
\affiliation{Department of Physics, University of Hamburg, D-20355 Hamburg, Germany}

\author{Robin Santra}
\email{robin.santra@cfel.de}
\affiliation{Center for Free-Electron Laser Science, DESY, D-22607 Hamburg, Germany}
\affiliation{Department of Physics, University of Hamburg, D-20355 Hamburg, Germany}


\begin{abstract}
 We present the extension of the time-dependent configuration interaction singles (TDCIS) method to the computation of the electron kinetic-energy spectrum in photoionization processes. Especially for strong and long ionizing light pulses the detection of the photoelectron poses a computational challenge because propagating the outgoing photoelectron wavepacket requires large grid sizes. Two different methods which allow for the extraction of the asymptotic photoelectron momentum are compared regarding their methodological and computational performance. The first method follows the scheme of Tong et al. \cite{tong} where the photoelectron wavefunction is absorbed by a real splitting function. The second method after Tao and Scrinzi \cite{scrinzi} measures the flux of the electron wavepacket through a surface at a fixed radius. With both methods the full angle- and energy-resolved photoelectron spectrum is obtained. Combined with the TDCIS scheme it is possible to analyze the dynamics of the outgoing electron in a channel-resolved way and, additionally, to study the dynamics of the bound electrons in the parent ion. As an application, one-photon and above-threshold ionization (ATI) of argon following strong XUV irradiation are studied via energy- and angle-resolved photoelectron spectra. 
\end{abstract}

\maketitle

\section{Introduction}\label{intro}

With the development of new light sources such as free-electron lasers (FELs) and attosecond laser sources the interest in strong-field physics and multiphoton processes has grown, because they provide the experimental means to control and image atomic and molecular systems and to test theoretical predictions of nonlinear processes \cite{corkumkrausz,chapman,levesque,spanner,stefanreview}. The photon energies of FELs extend from the UV to the X-ray range, and the intensities are such that they permit the investigation and control of inner-shell processes, Auger decay or above-threshold ionization (ATI) \cite{kanter,demekhin}. Above-threshold ionization, first observed in the 1970s \cite{ago}, is a highly nonlinear phenomenon in which more photons are absorbed than are needed for ionization. The pulse durations of FELs can be as short as a few femtoseconds \cite{emma}. With these pulse properties typical atomic timescales which extend from a few attoseconds to tens of femtoseconds can be accessed in order to study electronic dynamics in atoms, molecules and clusters \cite{kra,baker,wabnitz}.
In the strong-field regime, multiphoton processes play a significant role, especially if the photon energies lie in the UV to X-ray range \cite{young,moshammer,arina}. In general, in this frequency range a diversity of processes must be faced. The removal of a deep inner-shell electron is followed by various processes depending on the atomic states and the photon energy \cite{drescher}. If the laser pulse is strong enough, also multiphoton inner-shell ionization \cite{fukuzawa} as well as ATI processes can occur.

Experimentally, photoelectron spectroscopy is a powerful tool to analyze and quantify the processes that happen due to the irradiation of complex systems \cite{huefner,stolow}. For instance, in early experiments with intense light sources in the 1980s the angular distribution in ATI of xenon was measured \cite{fabre} in order to understand the ATI phenomenon. Synchrotron radiation was used to obtain high-quality angular distributions of electrons in photoionization of atoms \cite{schmidt,uwebecker}. Also in recent experiments photoelectron spectroscopy has been used to reveal decay mechanisms and multiphoton excitations in deep shells of atoms \cite{meyerpapa} and to understand the origin of the low-energy structure in strong-field ionization \cite{blaga,quan}. 

The process of photoionization has been studied extensively \cite{bethe,atomicphoto,krause,joachain,chelk}, e.g., in argon or xenon \cite{starace,hugovan}. In the weak-field limit, where the light-matter interaction can be treated perturbatively, the photoelectron spectrum has been calculated with methods that also include correlation effects. Most prominent examples are post-Hartree-Fock methods that use reference states, e.g. correlation methods like the configuration interaction \cite{shavitt,atomicphoto}, the coupled clusters method \cite{kuemmel} and the random-phase approximation \cite{amusia,dahl,delay}. Furthermore, approaches constructing continuum wavefunctions, like R-matrix theory \cite{burke}, have been applied to calculate photoionization cross sections \cite{burkelate} and photoelectron angular distributions \cite{taylor}; taking into account the interaction of the liberated electron with other atomic orbitals has led to the explanation of the giant dipole resonance of the $4d$ subshell in xenon \cite{atomicphoto}. 

In the strong-field regime the description of the ionized wavepacket is challenging due to the nonperturbative interaction between the electrons and the light pulse. Therefore, the calculation of photoelectron spectra is numerically more demanding than in the weak-field limit. Furthermore, many-body processes are often neglected in the strong-field regime and single-active electron (SAE) approaches have become a standard tool \cite{schaferkul,tong,bauer,popruzhenko} where correlation effects are omitted. Nevertheless, recently, extensions to many-body dynamics have been presented, e.g., R-matrix theory \cite{burkeburke,vander,lysa,timedelay,torlina2}, two-active electron \cite{djiokap,tarana,burgdoerfer} and time-dependent restricted-active-space configuration interaction theory \cite{hochstuhl,miyagi}.

Generally, the calculation of the photoelectron spectrum can be done after the pulse is over by projecting the photoelectron wavepacket  onto the eigenstates of the field-free continuum. However, this approach requires large numerical grids and its application is very limited even in the SAE cases. For this reason, new methods were developed to calculate the spectrum using wavepacket information in a fixed spatial volume much smaller than the volume that would be needed to fully encapsulate the wavepacket at the end of the strong-field pulse. There exist several approaches to overcome the obstacle of large grids, e.g., by measuring the electronic flux through a sphere at a fixed radius \cite{feuer} or splitting the wavefunction into an internal and an asymptotic part \cite{kellera,henkel} where the latter is then analyzed to yield the spectrum. The first implementation of the flux method in the strong-field case is the time-dependent surface flux (``tsurff'') method introduced by Tao and Scrinzi \cite{scrinzi}. It has recently been extended to the description of dissociation in molecules \cite{yue}. Tong et al. \cite{tong} applied the splitting approach to strong-field scenarios. With both methods double-differential photoelectron spectra can be calculated.

Our method for treating the electron dynamics within atoms is based on the time-dependent configuration interaction singles (TDCIS) scheme \cite{pab}. The Schr\"odinger equation is solved exactly by wavepacket propagation in the configuration interaction singles (CIS) basis. The TDCIS approach \cite{progxcid} includes interchannel coupling and allows investigating the wavepacket dynamics and, in particular, the impact of correlation effects between the photoelectron wavepacket and the remaining ion as discussed e.g. in Refs.~\cite{decoherence,giant,coopermin}. It is versatile with respect to the electric field properties (also multiple pulses can be chosen) and it has proven especially successful for strong-field studies \cite{giant,coopermin,transient,antonia1}.

In Sec.~\ref{theory}, we present the theoretical details of how the photoelectron spectrum is obtained within the TDCIS scheme: The wavefunction splitting method \cite{tong} is described in Sec.~\ref{splmethod} and the time-dependent surface flux method \cite{scrinzi} in Sec.~\ref{tsurff}. In Sec.~\ref{app} the two methods are analyzed with respect to their efficiency within TDCIS and are compared briefly. As an application we calculate and study the angle- and energy-resolved photoelectron spectrum of argon irradiated by strong XUV radiation. A summary and short outlook in Sec.~\ref{conc} conclude the article. Atomic units are used throughout except otherwise indicated.

\section{Theory}\label{theory}

\subsection{Theoretical background}
The time dependent Schr\"odinger equation of an $N$-electron system is given by 
\begin{equation}
 i \fr{\pa}{\pa t} |\Y^N(t)\rangle = \hat {H} (t) |\Y^N(t)\rangle . \label{schr}
\end{equation}
Considering linearly polarized light, the Hamiltonian takes the form 
\begin{equation}
 \hat{H}(t)= \hat {H}_0 +\hat{H}_1+\hat{\vec{p}} \cdot \vec{A}(t), \label{ham}
\end{equation}
where $\vec{A}(t)$ is the vector potential\footnote{Unlike in previous work on the TDCIS method \cite{roh,pab}, we use the velocity form at this point. Furthermore, the charge of the electron is negative, $q_e=-1$, so that $|q_e|=1$.}. Here, $\hat{H}(t)$ is the full $N$-electron Hamiltonian, $\hat{H}_0=\hat{T}+\hat{V}_{\rm nuc}+\hat{V}_{\rm MF}-E_{\rm HF}$ contains the kinetic energy $\hat{T}$, the nuclear potential $\hat{V}_{\rm nuc}$, the potential at the mean-field level $\hat{V}_{\rm MF}$ and the Hartree-Fock energy $E_{\rm HF}$, $\hat{H}_1=\frac{1}{|r_{12}|}-\hat{V}_{\rm MF}$ describes the Coulomb interactions beyond the mean-field level, and $\hat{\vec{p}}\cdot\vec{A}(t)$ is the light-matter interaction within the velocity form in the dipole approximation.

Within the CIS approach only one-particle--one-hole excitations $|\Phi_i^a\rangle$ with respect to the Hartree-Fock ground state $|\Phi_0\rangle$ are considered. Therefore, the wavefunction (now omitting the superscript $N$) is expanded in the CIS basis as
\begin{align}
 |\Psi(t)\rangle&=\alpha_0(t)|\Phi_0\rangle+\sum_{i,a}\alpha_i^a(t)|\Phi_i^a\rangle,\label{tdcis}
\end{align}
where the index $i$ symbolizes an initially occupied orbital and $a$ denotes an unoccupied (virtual) orbital to which the particle can be excited: $|\Phi_i^a\rangle=\fr{1}{\sqrt{2}}\left( \hat{c}_{a+}^\dagger \hat{c}_{i+}+ \hat{c}_{a-}^\dagger \hat{c}_{i-} \right) |\Phi_0\rangle$. The operators $\hat{c}_{p\s}^\dagger$ and $\hat{c}_{p\s}$ create and annihilate electrons, respectively, in the spin orbitals $|\varphi _{p\s}\rangle$. The total spin is not altered in the considered processes ($S=0$), so that only spin singlets occur. Therefore, and for the sake of readability, we drop the spin index and treat the spatial part of the orbitals $|\varphi _p\rangle$. Inserting the wavefunction expansion (\ref{tdcis}) into the Schr\"odinger equation (\ref{schr}) and projecting onto the states $|\Phi_0\rangle$ and $|\Phi_i^a\rangle$ yields the following equations of motion for the expansion coefficients $\a_i^a(t)$:
\begin{subequations}
\begin{align}
 i\dot{\a}_0(t) &= \vec{A}(t)\cdot\sum_{i,a}\langle \Phi_0 | \, \hat{\vec{p}}\, | \Phi_i^a\rangle \a_i^a(t),\label{cis1}\\
 i\dot{\a}_i^a(t) &= (\varepsilon_a-\varepsilon_i) \a_i^a(t) + \sum_{j,b} \langle \Phi_i^a |\hat{H}_1|\Phi_j^b    \rangle    \a_j^b(t) \nn\\
                 &  +\vec{A}(t)\cdot\left( \langle\Phi_i^a | \, \hat{\vec{p}}\, |\Phi_0  \rangle \a_0(t) + \sum_{j,b}\langle \Phi_i^a |\, \hat{\vec{p}}\,|\Phi_j^b    \rangle    \a_j^b(t) \right ),\label{cis2}
\end{align}
\end{subequations}
where $\varepsilon_p$ denotes the energy of the orbital $|\varphi_p\rangle$ ($\hat{H}_0|\varphi_p\rangle = \varepsilon_p |\varphi_p\rangle$). As introduced in Ref.~\cite{roh}, for each ionization channel all single excitations from the occupied orbital $|\varphi _i\rangle$ may be collected in one ``channel wavefunction'':
\begin{equation}
 |\chi_i(t)\rangle = \sum_a \alpha_i^a(t)|\varphi_a \rangle. \label{channelwfct}
\end{equation}
These channel wavefunctions may now be used to calculate all quantities in a channel-resolved manner. In this way, effectively one-particle wavefunctions are obtained, which will be used in the following to derive the formulae for the photoelectron spectra. A detailed description of the TDCIS method can be found in Refs.~\cite{pab,roh}. The TDCIS method provides the coefficients of the wavefunction in the CIS basis, which are propagated in time. During the propagation, quantities that are needed for the calculation of the photoelectron spectrum are prepared using the channel wavefunction coefficients. After the propagation, these quantities are then used in the subsequent analysis step to determine the spectral components of the channel wavefunctions. At the end, an incoherent summation over all ionization channels is performed to obtain the photoelectron spectrum. The two analysis methods are described in the following.

\subsection{Wavefunction splitting method}\label{splmethod}

We describe the concrete implementation of the splitting method introduced by Tong et al. in Ref.~\cite{tong} within our time-dependent propagation scheme. A real radial splitting function of the form
\begin{equation}
 \hat{S}= \left[1+e^{-(\hat{r}-r_c)/\D}\right]^{-1}
\end{equation}
is used to smoothly split the channel wavefunction~\eqref{channelwfct}. The parameter $r_c$ denotes the radius where the splitting function is centered, and $\D$ is a ``smoothing'' parameter controlling the slope of the function. At the first splitting time step $t_0$ the channel wavefunction is split into two parts (for each channel $i$):
\begin{equation}
 |\chi_i(t_0)\rangle = (1-\hat{S})|\chi_i(t_0)\rangle + \hat{S} |\chi_i(t_0)\rangle \equiv |\chi_{i, \rm in}(t_0)\rangle + |\chi_{i, \rm out}(t_0)\rangle.
\end{equation}
$|\chi_{i,\rm in}(t)\rangle$ is the wavefunction in the inner region $0<r \lesssim  r_c$ and $|\chi_{i,\rm out}(t)\rangle$ is the wavefunction in the outer region $r_c\lesssim r\le r_{\rm max} $. 
Then, the following procedure is performed at $t_0$: The outer part of the wavefunction $|\chi_{i,\rm out}(t_0)\rangle$ is analytically propagated to a long time $T$ after the laser pulse is over using the Volkov Hamiltonian $\hat{H}_V(\tau)$ with the time propagator 
\begin{equation}
 \hat{ U}_V(t_2,t_1)=\exp\left(-i \int_{t_1}^{t_2}  \hat{H}_V(\tau) d\t \right),\ \ \ \ \hat{H}_V(\tau)=\frac{1}{2} \left[ \hat{\vec{p}}+\vec{A}(\t) \right]^2,
\end{equation}
 under the assumption that far from the atom the electron experiences only the laser field and not the Coulomb field of the parent ion. It is also assumed that, at the splitting radius, the electron is sufficiently far away to not return to the ion. 

The inner part of the wavefunction $|\chi_{i, \rm in}(t_0)\rangle$ is propagated on a numerical grid using the full CIS Hamiltonian [see Eqs.~\eqref{cis1} and \eqref{cis2}]. For the splitting function the ratio $r_c/\Delta\gg 1$ must be chosen such that the ground state $|\Phi_0\rangle$ is not affected by the splitting: $\hat{S}|\Phi_0\rangle=0$. 

At the next splitting time $t_1$ the inner part of the wavefunction which was propagated from $t_0$ to $t_1$ is split again. Thus, the following prescription is obtained:
\begin{align}
 |\chi_{i, \rm in}(t_j)\rangle \rightarrow  |\tilde{\chi}_{i}(t_{j+1})\rangle = |\chi_{i, \rm in}(t_{j+1})\rangle +|\chi_{i, \rm out}(t_{j+1})\rangle.
\end{align}
This is now repeated for every splitting time $t_j$, until all parts of the electron wavepacket that are of interest have reached the outer region. Each $|\chi_{i, \rm out}(t_{j+1})\rangle$ is again propagated analytically to $t=T$.

Computationally, $|\chi_{i,\rm out}(t_j)\rangle$ is initially expressed in the CIS basis. For this purpose, we define new expansion coefficients for the outer wavefunction
\begin{equation} 
 \beta_i^a(t_j)=\langle \varphi_a|\hat{S}|\tilde{\chi}_i(t_j)\rangle,\label{newcoeff}
\end{equation}
and express the wavefunction in the outer region as $|\chi_{i,\rm out}(t_j)\rangle=\sum_a\beta_i^a(t_j)|\varphi_a\rangle$. 
 During the propagation, at every splitting time step $t_j$, which can be ---and for computational efficiency should be--- a multiple of the actual propagation time step, the splitting function $\hat{S}$ is applied and the expansion coefficients \eqref{newcoeff} are calculated and stored. Since the outer wavefunction is split from the inner part and treated analytically, the grid size needed for the description of the wavefunction is automatically reduced. Later, when the spectrum is calculated, the coefficients $\beta_i^a$ are inserted and used for the analysis. 

The Volkov states $| \Y_{\vec {p}}^{V}\rangle\equiv| \vec{p}\,^V\rangle$ are eigenstates of the Volkov Hamiltonian and form a basis set in which the channel wavepacket at time $T$ can be expanded:
\begin{equation}
 |\chi_{i,\rm out}(T)\rangle = \int\! d^3p \sum_{t_j}C_{i}(\vec{p},t_j) \,| \vec{p}\,^V\rangle\equiv\int \!d^3p\ \tilde{C}_i(\vec{p}\,)\,| \vec{p}\,^V\rangle. \label{coeff}
\end{equation}
In the velocity form the Volkov states are nothing but plane waves $\Y_{\vec {p}}^{V}(\vec{r})= (2\p)^{-3/2}e^{i\vec{p}\cdot\vec{r}}$.
The photoelectron spectrum is obtained by calculating the spectral components of the outer wavefunction. For this purpose, the following coefficients are evaluated:
\begin{equation}
 C_i(\vec{p},t_j)=\int d^3p'\langle  \vec{p}\,^V|\hat{ U}_V(T,t_j)| \vec{p}\,'\,\!^V\rangle \underbrace{\langle \vec{p}\,'\,\!^V | \chi_{i,\rm out}(t_j)  \rangle}_{c_i(\vec {p}\,',t_j)}.
\end{equation}
First, we calculate the $c_i(\vec{p},t_j)$ for each splitting time $t_j$
\begin{equation}
 c_i(\vec{p},t_j)=(2\p)^{-3/2}\sum_a \beta_i^a(t_j)\!\int d^3 r e^{-i\vec{p}\cdot \vec{r}} \varphi_a(\vec{r}),\label{expcoeff}
\end{equation}
where the orbital is now explicitly given in the spatial representation by $\langle \vec{r}|\hat{c}_a^\dagger|0 \rangle=\langle \vec{r}|\varphi_a\rangle= \varphi_a(\vec{r})=\fr{u_{n_a,l_a}(r)}{r}Y_{l_a,m_a}(\W_{\vec{r}})$ and, thus, possesses a radial and an angular part. 
We use the multipole expansion for the exponential function
\begin{equation}
 e^{i\vec{p}\cdot\vec{r}}=4\p\sum_{l=0}^\infty i^l  j_l(pr)\sum_{m=-l}^l Y^*_{lm}(\W_{\vec{p}})Y_{lm}(\W_{\vec{r}}),\label{multipole}
\end{equation}
where $j_l(pr)$ denotes the spherical Bessel function of order $l$. The orthonormality relations of the spherical harmonics reduce the three-dimensional integrals in Eq.~\eqref{expcoeff} to one-dimensional radial integrals. Finally, propagating to a long time $T$ after the pulse, we obtain the coefficients
\begin{align}
 C_i(\vec{p},t_j)&=\langle  \vec{p}\,^V|\hat{ U}_V(T,t_j)| \chi_{i,\rm out}(t_j)  \rangle\label{finalcoeff} \\
&= \sqrt{\fr{2}{\p}}\exp\left( -\frac{i}{2} \int_{t_j}^{T}\!\!d\t\left[\vec{p}\!+\!\vec{A}(\t) \right]^2 \right) \sum_a (-i)^{l_a}\beta_i^a(t_j)Y_{l_a,m_a}\!(\W_{\vec{p}})\int\! dr\thinspace r\thinspace u_{n_a,l_a}\!(r)j_{l_a}\!(pr) . \nn
\end{align}
These coefficients can be used to calculate the angle and energy distribution of the ejected electron because at time $T$ the canonical momentum equals the kinetic momentum. One can choose now a homogeneous momentum grid and calculate these coefficients for each splitting time step.
In order to obtain the full electron wavepacket at time $T$ all contributions from splitting times $t_j$ must be summed up coherently to obtain the coefficients $\tilde{C}_i(\vec{p}\,)$ in Eq.~\eqref{coeff} for each ionization channel $i$. Then, incoherent summation over all possible ionization channels yields the photoelectron spectrum:
\begin{equation}
 \fr{d^2P(\vec{p})}{dE d\W}=p \sum_i \big| \tilde{ C}_i(\vec{p}\, ) \big|^2.
\end{equation}
The extra factor of $p$ results from the conversion from the momentum to the energy differential. As long as the time $T$ is chosen to be after the pulse the result is $T$ independent. Of course, one needs to choose a sufficiently large $T$ such that the parts of the electron wavefunction that one wants to record have entered the outer region and can be analyzed.

\subsection{Time-dependent surface flux method (``tsurff'')}\label{tsurff}

The second method for the calculation of photoelectron spectra is based on the approach presented by Tao and Scrinzi in Ref.~\cite{scrinzi} where it was used to calculate strong-field infrared photoionization spectra in combination with infinite-range exterior complex scaling \cite{perfectabs}. In this approach the electron wavefunction is analyzed during its evolution when crossing the surface of a sphere of a given radius $r_c$. Again, it is assumed that the wavefunction can be split into two parts: One part is bound to the atom and is a solution to the full Hamiltonian, the other part can be viewed as free from the parent ion and is a solution to the Volkov Hamiltonian. Therefore, the method also relies conceptually on a splitting procedure. Nevertheless, and in contrast to the splitting method, the wavefunction is not altered in this process. As above, the key idea is to obtain the spectral components of the wavefunction by projecting onto plane waves. 

The surface radius $r_c$ is chosen such that the electron can be considered to be free, and a sufficiently large time $T$ after the pulse is over is picked by which the electron with the kinetic energy of interest has passed this surface. (For very low-energy electrons a correspondingly larger time has to be chosen.) At this time, the channel wavefunction $|\chi_i\rangle$ for each ionization channel $i$ can be split into a bound part (corresponding to the inner wavefunction in the splitting method) and an asymptotic part, which describes the ionized contribution: $|\chi_i(T)\rangle=|\chi_{i, \rm in}(T)\rangle+|\chi_{i ,\rm out}(T)\rangle$. As in Sec.~\ref{splmethod}, the system Hamiltonian for distances larger than $r_c$ is approximated by the Volkov Hamiltonian. Using the Volkov states of Sec.~\ref{splmethod} and $|\Y_{\vec {p}}^{V}(T)\rangle=\hat{U}_V(T,-\infty)|\Y_{\vec {p}}^{V}\rangle$, the outer wavefunction is represented as follows:
\begin{equation}
 |\chi_{i,\rm out}(T)\rangle=\int\! d^3 p\  b_i(\vec{p})\  |\Y_{\vec {p}}^{V}(T)\rangle,
\end{equation}
which vanishes for $r\le r_c$. Thus, the photoelectron spectrum is the sum, over all channels, of the $|b_i(\vec{p})|^2$, where
\begin{equation}
 |b_i(\vec{p})|^2 =\bigg| \int_{r>r_c}\!\!\!\!\!\!\! d^3r\  \Y_{\vec {p}}^{V*}(\vec{r},T)\ \chi_{i,\rm out}(\vec{r},T) \bigg|^2=: \big|\langle \Y_{\vec {p}}^{V}(T)|\theta(\hat{r}-r_c)|\chi_{i,\rm out}(T)\rangle \big|^2.
\end{equation}
Here, the Heaviside step function $\theta$ enters (we adopted the notation by Tao and Scrinzi \cite{scrinzi}). In order to avoid the need for a representation of $\chi_{i,\rm out}(\vec{r},T)$ at large $r$ (because $T$ is large, a fast electron moves far out during this time), this $3D$-integral is converted into a time integral involving the wavefunction only at $r=r_c$. For that, we must know the time evolution of the asymptotic part of the wavefunction after it has passed the surface. Inserting the Schr\"odinger equation where necessary and using the Volkov solutions in the velocity form outside the sphere with radius $r_c$ we obtain \cite{scrinzi}
\begin{equation}
 \langle \Y_{\vec {p}}^{V}(T)|\theta(\hat{r}-r_c) |\chi_{i,\rm out}(T)\rangle = i\int_{-\infty}^T\!\!\! dt\thinspace  \langle\Y_{\vec {p}}^{V}(t)|\left[ -\fr{1}{2}\D -i \vec{A} (t)\cdot \vec{\nabla}, \theta(\hat{r}-r_c)  \right]|\chi_{i,\rm out }(t)\rangle.\label{exprtsurff}
\end{equation}
 The commutator, which vanishes everywhere except at $r=r_c$, is easily evaluated in polar coordinates (assuming linear polarization) and we obtain
\begin{equation}
 \left[ -\fr{1}{2}\D -i \vec{A} (t)\cdot \vec{\nabla}, \theta(\hat{r}-r_c)  \right] = -\fr{1}{2r^2} \pa_r r^2 \d(r-r_c) -\fr{1}{2}\d(r-r_c) \pa_r+iA(t)\cos(\theta) \d(r-r_c).
\end{equation}
More details can be found in Ref.~\cite{scrinzi} as well as in Ref.~\cite{tsurff}. We shuffle the derivative in the first operator term to the left, via integration by parts, and obtain the operator
\begin{equation}
 -\fr{1}{r} \d(r-r_c) +\cev{\pa_r} \fr{1}{2}\d(r-r_c) -\fr{1}{2}\d(r-r_c) \vec{\pa}_r+iA(t)\cos(\theta) \d(r-r_c),
\end{equation}
where ``$\cev{\pa} $'' means that the derivative acts to the left on the Volkov state and ``$ \vec{\pa} $'' means that the derivative acts to the right on the channel wavefunction. In order to implement this operator acting on the channel wavefunctions we have to calculate also the first derivative of the wavefunctions with respect to $r$ at the radius $r_c$. After the propagation, during which the coefficients of the channel wavefunctions $\chi_i(r_c,t)$ as well as of their first derivatives $[\pa_r\chi_i(r,t)|_{r=r_c}]$ have been calculated, the expression \eqref{exprtsurff} can be computed. Since we introduce the multipole expansion [see Eq.~\eqref{multipole}] for the Volkov states we have to calculate also derivatives of the spherical Bessel functions at the radius $r_c$. This calculation is performed during the analysis step for each angular momentum $l$. In the last term we express the cosine as a spherical harmonic and use the identity for the integral over three spherical harmonics
\[\int  d\W \  Y_{l_3,m_3}^*\!(\W)Y_{l_2,m_2}\!(\W) Y_{l_1, m_1}\! (\W)\!=\! \fr{\sqrt{(2l_1\!+\!1)(2l_2\!+\!1)}}{4\p(2l_3\!+\!1)}C^{l_3m_3}_{l_1m_1,l_2m_2}C^{l_30}_{l_10,l_20}, \]
where the Clebsch-Gordan coefficients are given by $C^{l_3m_3}_{l_1m_1,l_2m_2}=\langle l_1m_1,l_2m_2| l_3m_3 \rangle$. Thus, we obtain the spectral components in their final form:
\begin{align}
 \langle \Y_{\vec {p}}^{V}(T)&|\theta(\hat{r}-r_c)|\chi_{i,\rm out}(T)\rangle = i \sqrt{\fr{2}{\p}}  \int_{-\infty}^T\!\!\! dt\ \!\exp\left( -\frac{i}{2} \int_{-\infty}^{t}\!\!\!d\t\left[\vec{p}+\vec{A}(\t) \right]^2 \right)\sum_{a}\\ 
&\left\{ (-i)^{l_a}\left[-j_{l_a}\!(pr_c)+ \fr{pr_c}{2}\,j'_{l_a}(pr_c) -\fr{1}{2} j_{l_a}\!(pr_c)  \right] Y_{l_a,m_a}\!(\W_p)\,u_{n_a,l_a}\!(r_c)\,\alpha_i^a(t)\right.\nn\\
&\left.-\fr{(-i)^{l_a}}{2} j_{l_a}\!(pr_c)\,Y_{l_a,m_a}\!(\W_p)\, u'_{n_a,l_a}\!(r_c)
\,\alpha_i^a(t)\right.\nn\\
&\left.+\fr{i}{2\sqrt{\p}}\,r_c\, u_{n_a,l_a}\!(r_c)\,A(t)\,\alpha_i^a(t) \sum_{l=0}^\infty (-i)^lj_l(pr_c)\fr{\sqrt{2l_a\!+\!1}}{2l+1} C^{l,m_a}_{l_a,m_a;1,0} C^{l,0}_{l_a,0;1,0}   Y_{l,m_a}\!(\W_p)\right\}, \nn
\end{align}
where $j'_{l_a}(pr_c)=\pa_z j_{l_a}\!(z)\big|_{z=pr_c}$ and $u'_{n_a,l_a}\!(r_c)=\pa_ru_{n_a,l_a}\!(r)\big|_{r=r_c}$. Although this expression may seem fairly complicated, it involves only quantities evaluated at one single radius $r=r_c$.
The photoelectron spectrum is then obtained as
\begin{equation}
 \fr{d^2P(\vec{p})}{dE d\W}= p\sum_i \big| \langle \Y_{\vec {p}}^{V}(T)|\theta(\hat{r}-r_c)|\chi_{i,\rm out}(T)\rangle\big| ^2,
\end{equation}
where, as in the splitting method, the incoherent sum over all ionization channels $i$ is performed. In the present implementation, a complex absorbing potential (CAP) absorbs the wavefunction near the end of the numerical grid \cite{rismey,cedersantra,muga}. We use a CAP of the form $W(r)=\theta(r-r_{\rm CAP})(r-r_{\rm CAP})^2$, where $\theta$ is again the Heaviside step function and $r_{\rm CAP}$ is the radius where the CAP starts absorbing. It is added to the Hamiltonian in Eq.~\eqref{ham} in the form $-i\,\h\, \hat{W}$, where $\h$ is the CAP strength. As will be discussed in Sec.~\ref{tsurfapp} the absorption via a CAP has to be optimized carefully, because reflections from the end of the numerical grid as well as from the CAP itself have to be minimized in order to obtain an accurate photoelectron spectrum.

\section{Application: Argon under strong XUV radiation}\label{app}

With two methods for the calculation of photoelectron spectra implemented in TDCIS, we investigate one-photon and above-threshold ionization processes of argon in the XUV regime. Motivated by a recent experiment carried out at the free-electron laser facility FLASH in Hamburg \cite{meyerexp} we assume a photon energy of $105$~eV, which is far above the threshold for the ionization out of the $3$p and $3$s subshells. In the following we examine the functionality of the splitting and the surface flux methods by means of the specific example of ionization of argon in the XUV. 

\subsection{Wavefunction splitting method}\label{wsplmethod}
In the splitting method, three parameters have to be adjusted: the splitting radius, the smoothness of the splitting function and the rate at which the absorption is applied. A first criterion for verifying that the absorption through the masking function is performed correctly is the comparison of the total ground state population obtained via splitting with the population obtained with the CAP. We use a Gaussian pulse with $9\times 10^{13}$~Wcm$^{-2}$ peak intensity and $1.2$~fs duration (full width at half maximum, FWHM) at a photon energy of $105$~eV. For this pulse a converged result for the CAP strength $\h=1\times10^{-3}$, $r_{\rm max}=150$~a.u., and $r_{\rm max}\!-\!r_{\rm CAP}=30$~a.u. gives an ionization probability of $7.197\times 10^{-3}$ after the pulse. In the studied parameter cases the agreement between the splitting results and that CAP result is better than $3\times 10^{-3}$ relative difference (choosing, e.g., $r_{\rm max}=150$~a.u., $r_c=80$~a.u., $\D=10$~a.u., and varying the splitting time step between $0.2$~a.u. and $10$~a.u.). With more frequent absorption the agreement gets slightly better.

\begin{figure}[htbp]
 \centering
 \includegraphics[width=10cm]{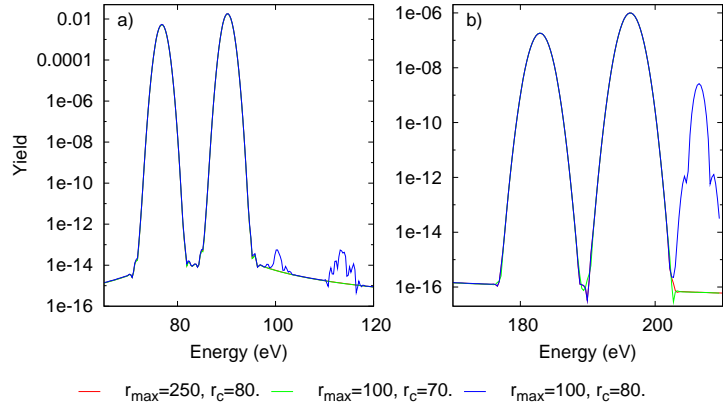}
 \caption{The photoelectron spectrum of argon for a pulse with $105$~eV photon energy, $9\times 10^{13}$~Wcm$^{-2}$ intensity and $1.2$~fs duration is shown for different radial grid sizes $r_{\rm max}$ and splitting radii $r_c$. The smoothing parameter is $\D=10$~a.u. and the splitting time step is ${\rm d }t_{\rm spl}=0.2$~a.u. Panel a) shows the one-photon absorption lines, panel b) shows the energetically lowest ATI lines for different splitting radii. All radii are given in atomic units. The spectrum does not change under variation of the splitting radius as long as around $30$~a.u. units are left for absorption.}
 \label{fig1}
\end{figure}

 Analyzing the splitting method, we find that the splitting radius $r_c$ and the smoothing parameter $\D$ can be varied rather freely without changing the (physical) spectrum. Although the total radial grid size can be chosen as small as $100$~a.u. (cf. Fig.~\ref{fig1}) we choose also a larger radial grid extension with $r_{\rm max}~=250$~a.u. and vary the splitting radius in the wide range from 80~a.u. to 230~a.u. Exemplarily,  results in the direction $\theta=0$ (along the XUV polarization axis) when the radial grid size and splitting radius are varied are shown in Fig.~\ref{fig1}(a). The spectrum shows the one-photon absorption peaks at the energy corresponding to the difference between photon energy and binding energy of the corresponding orbital ($3s$ and $3p$, respectively). The second part of the spectrum, in Fig.~\ref{fig1}(b), is separated from the first part by the photon energy and is, therefore, attributed to above-threshold ionization. The width of the peaks corresponds to the Fourier-limited energy width according to $\t\D \w= 2.765$ (all quantities in atomic units), where $\t$ is the duration of the pulse intensity envelope (FWHM) and $\D\w$ is the bandwidth of the power spectrum (FWHM). The figure shows that the spectrum is independent of the splitting radius as long as around 30~a.u. are left to the end of the numerical grid for absorption. Reducing the difference $r_{\rm max}-r_c$ to 20~a.u. produces artificial peaks near the physical peaks. To estimate how large the absorption range must be let us consider an electron with $200$~eV kinetic energy. It covers a distance of roughly $4$~a.u. per atomic unit of time. The numerical results show that the range over which the wavefunction is absorbed by the splitting function must be much larger than this distance (almost 10 times larger) in order to avoid reflections. This can be understood if one considers that the slope of the splitting function at a smoothing parameter of $\D=10$~a.u. extends over a range of around $30$~a.u. beyond the splitting radius to reach $95\%$ absorption of the wavefunction.

\begin{figure}[htbp]
  \centering
  \includegraphics[width=10cm]{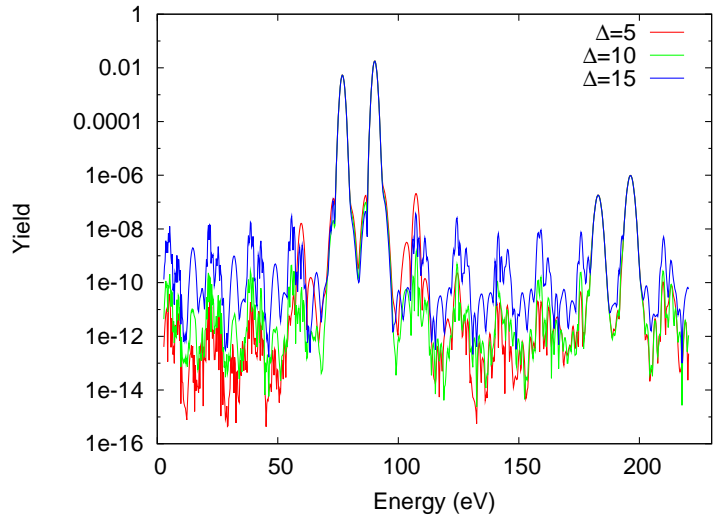}
  \caption{The argon photoelectron spectrum is shown for different smoothing parameters $\D$. The pulse parameters are the same as for Fig.~\ref{fig1}. The radial grid size is $r_{\rm max}=150$~a.u., the splitting radius is $80$~a.u., and the splitting is applied every $10$~a.u. of time. The 3p and 3s peaks are not affected by the change of the slope of the splitting function, although the numerical noise resulting from reflections from the splitting function changes.}
  \label{fig2}
 \end{figure}

\begin{figure}[htbp]
 \centering
 \includegraphics[width=10cm]{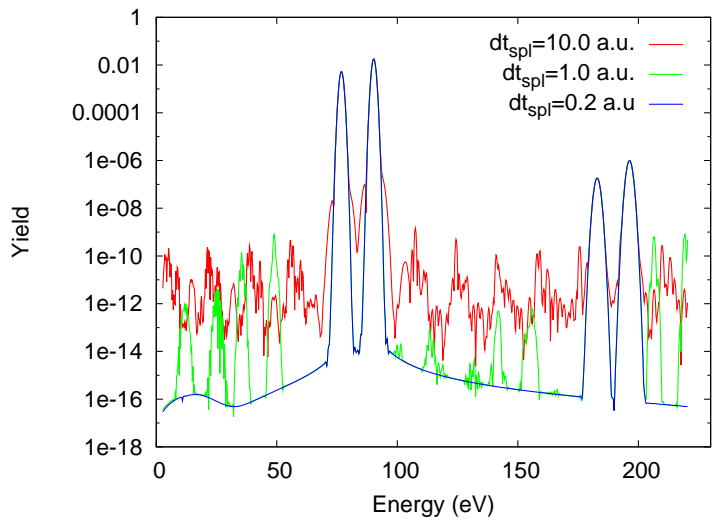}
 \caption{The argon photoelectron spectrum in the polarization direction is shown. The variation of the splitting time step results in significant changes in the (numerical) oscillations. The pulse parameters are the same as for Figs.~\ref{fig1} and \ref{fig2}. At a fixed smoothing parameter $\D=10$~a.u., $r_{\rm max}=150$~a.u., $r_c=80$~a.u., the noise is suppressed by several orders of magnitude for more frequent splitting. The one-photon peaks and ATI peaks do not change significantly.}
 \label{fig3}
\end{figure}

Since the splitting radius is not very crucial for the spectrum we proceed to the variation of the other parameters. Spectra for various smoothing parameters $\D$ are shown in Fig.~\ref{fig2}. Here, the radial grid size is kept fixed at $r_{\rm max}=150$~a.u., the splitting radius is $80$~a.u., and absorption is performed every $10$~a.u. of time. The physical peaks are reproduced correctly for all $\D$, the noise amplitude, however, is changing. A value of $\D=10$ seems to be the optimum, for $\D=15$ the amplitude of unphysical peaks is higher, while the steeper slope corresponding to $\D=5$~a.u. produces higher oscillations near the physical peaks, which should be avoided.
The method is particularly sensitive to the splitting rate, i.e., how often the splitting is applied. The more frequently the splitting function is applied the less noise is obtained. This is shown in Fig.~\ref{fig3}, where only the splitting time step is varied, while the radial grid size is kept constant at $150$~a.u., the splitting radius is set to $80$~a.u. and the smoothing parameter is $10$~a.u. We find that the unphysical peaks or artifacts do not contribute to the physical observables because they are orders of magnitude smaller. The noisy oscillations result from numerical issues, e.g., the higher the frequency of splitting the more reflections are accumulated from the slope of the splitting function. For this reason, the choice of the slope of the splitting function is coupled to the frequency of splitting. For more frequent absorption of the wavefunction the steepness should be reduced. Since for every splitting time step the new coefficients $\beta_i^a(t_j)$ have to be calculated and stored during the propagation and the quantities \eqref{finalcoeff} have to be evaluated during the analysis step, it is not convenient to perform the splitting at every propagation time step as mentioned in Sec.~\ref{splmethod}. In the calculations shown the propagation time step is $0.05$~a.u.

From the derivation of the splitting method in subsection \ref{splmethod} it can be seen that the electron spectrum is normalized to the total ionization probability (because only normalized wavefunctions are used). Therefore, the integrated spectrum represents a good measure of the quality of the spectrum; the fully integrated spectrum must agree with the total ionization probability. This can be verified for different parameter specifications. The relative difference to the CAP result is found to be smaller than $2\%$ in all studied parameter cases.

\begin{figure}[htbp]
 \centering
 \includegraphics[width=\linewidth]{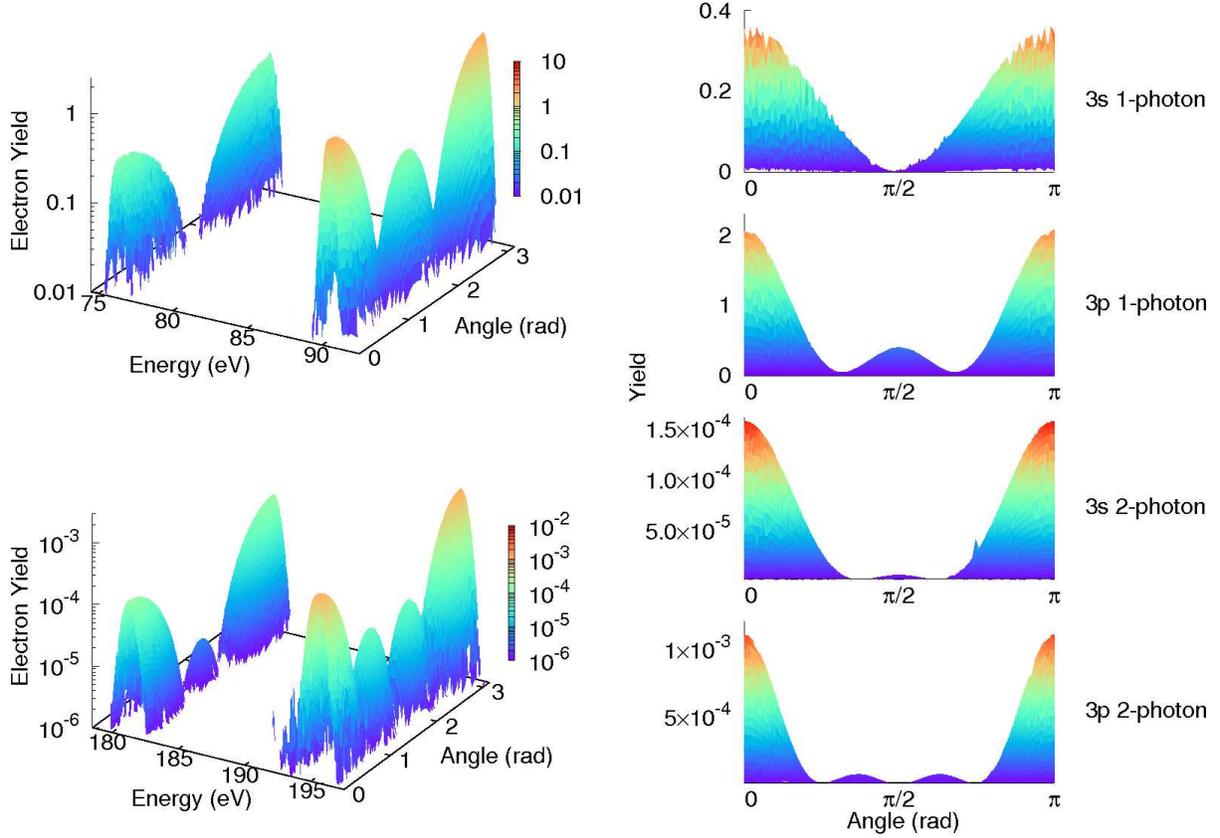}
 \caption{The energy- and angle-resolved argon photoelectron spectrum produced with the splitting method is shown for an XUV pulse at $105$~eV photon energy, $1.0\times 10^{15}$~Wcm$^{-2}$ intensity and $2.6$ fs pulse duration. The grid size is $r_{\rm max}=100$~a.u., $r_c=20$~a.u., $\D=5$~a.u., and d$t_{\rm spl}=10$~a.u. The angle denotes the direction with respect to the polarization axis of the pulse. The angular distribution reflects the change in angular momentum by multiphoton absorption.}
 \label{fig4}
\end{figure}

In the following we apply a strong XUV pulse centered at $105$~eV with $0.7$~eV bandwidth (FWHM), which corresponds to a Fourier-transform limited pulse with 108 a.u. ($2.6$ fs) duration. The peak intensity of the pulse is $1.0\times 10^{15}$~Wcm$^{-2}$. In the upper left panel of Fig.~\ref{fig4} the full angle- and energy-resolved photoelectron spectrum of argon after one-photon absorption is shown. The angle denotes the direction with respect to the polarization axis. The peaks arise from ionization out of the 3s and 3p shells, respectively. The lower left panel shows the corresponding ATI spectrum. As expected, the angular distributions feature the corresponding contributions from the different channels, which can be seen on the right in the four cuts along the fixed peak energies: The one-photon peak from the 3s shows a p-wave character, the 3p peak has both an s- and a d-wave contribution. Analogously, the two-photon peak of 3s exhibits an s- and d-wave character and the 3p peak a p- and f-wave character.
In a nutshell, the splitting method is a well-working tool for the calculation of photoelectron spectra although the requirement to optimize three parameters ($r_c,$ $\D$, d$t_{\rm spl}$) can render calculations time-consuming.

\subsection{Time-dependent surface flux method (tsurff)}\label{tsurfapp}
\begin{figure}[htbp]
 \centering
 \includegraphics[width=10cm]{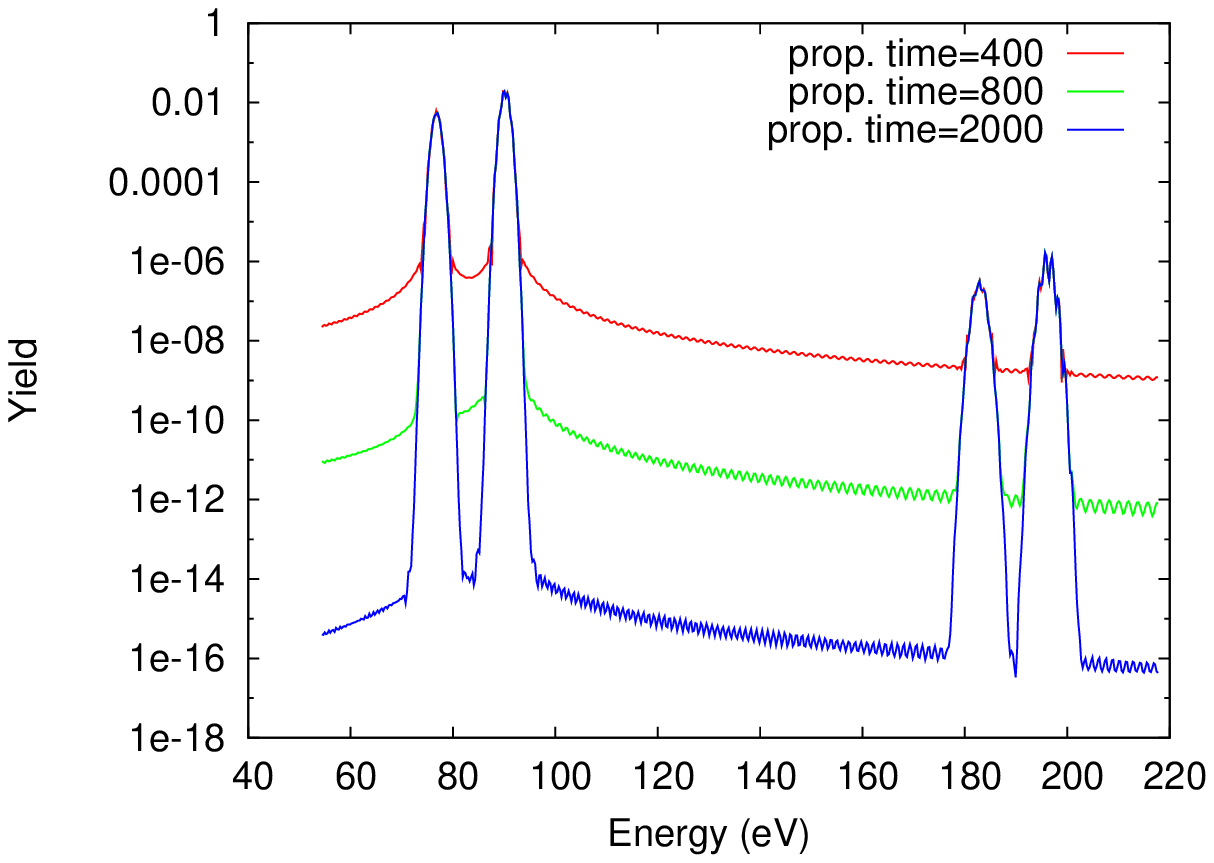}
  \caption{The argon photoelectron spectrum calculated via tsurff is shown for different propagation times (in a.u.). The pulse parameters are the same as in Figs.~\ref{fig1} to \ref{fig3}. The computational parameter specifications are: $r_{\rm max}=250$~a.u., $r_{\rm CAP}=230$~a.u., $\h=1\times10^{-3}$, and $r_c=180$~a.u. The oscillations decrease by orders of magnitude for longer propagation.}
 \label{figtime}
\end{figure}

We turn now to the tsurff method. The method depends on the radius $r_c$ where the surface measuring the flux is placed and on the parameters of the absorption method. As already mentioned in Sec.~\ref{tsurff}, in the present work the absorption is performed with a CAP, which depends on two parameters: the CAP strength $\h$ and the radius $r_{\rm CAP}$ where the CAP starts absorbing. For tsurff also the total propagation time plays an important role. While the splitting method is not affected by a variation of the propagation time (as long as it is longer than the pulse and long enough for the electronic wavepacket of interest to enter the absorption region), tsurff requires a long propagation. This is shown in Fig.~\ref{figtime}. The noise level decreases dramatically with longer time propagation.
On the other hand, the calculation of the spectrum itself can be performed faster than with the splitting method, because no radial integrals are involved. Instead, all quantities are evaluated at the radius $r=r_c$.

\begin{figure}[htbp]
 \centering
 \includegraphics[width=10cm]{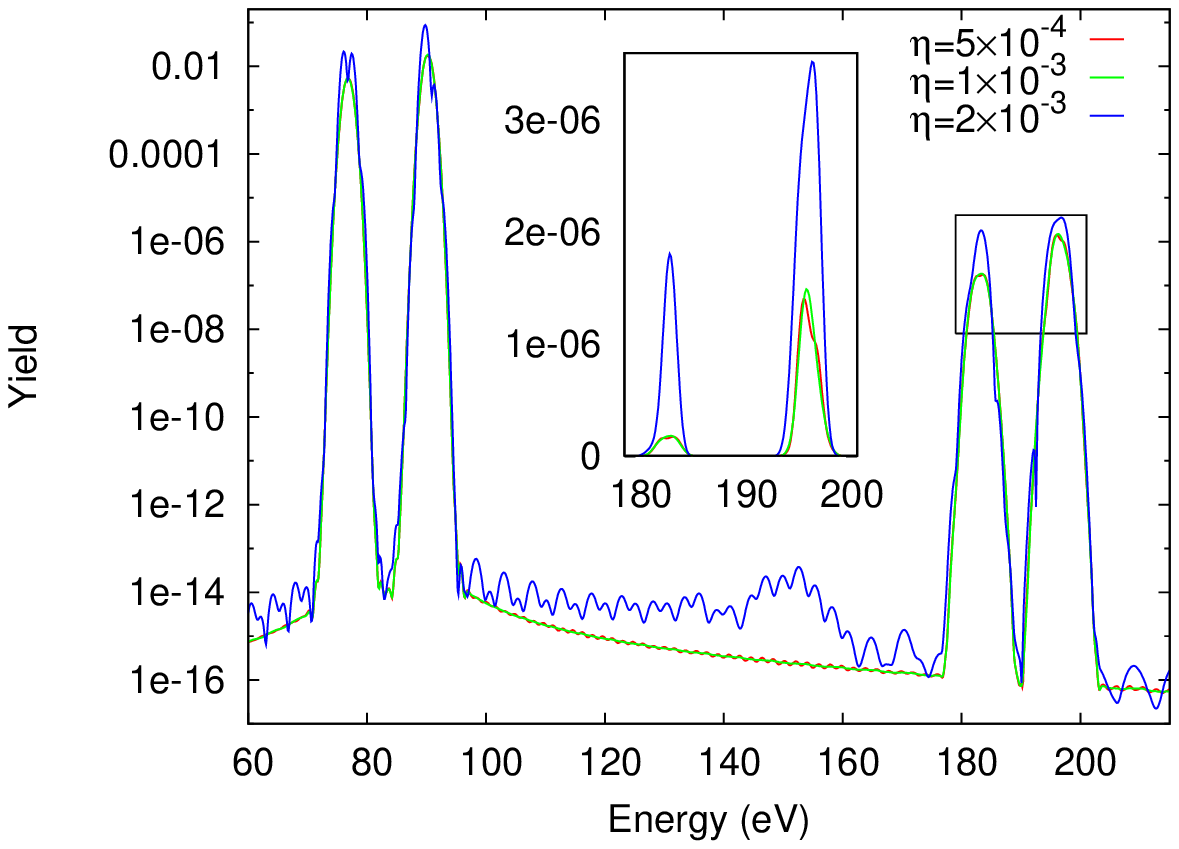}
  \caption{The argon photoelectron spectrum along the polarization axis of the field, calculated with tsurff, is shown for a pulse with $105$~eV photon energy, $9\times 10^{13}$~Wcm$^{-2}$ intensity and $1.2$~fs duration for different CAP strengths. The radial grid size is $r_{\rm max}=150$~a.u., the CAP radius is $r_{\rm CAP}=120$~a.u., and $r_c=100$~a.u. The oscillations are due to reflections from the end of the radial grid and/or the CAP.}
 \label{fig5}
\end{figure}

\begin{figure}[htbp]
 \centering
 \includegraphics[width=10cm]{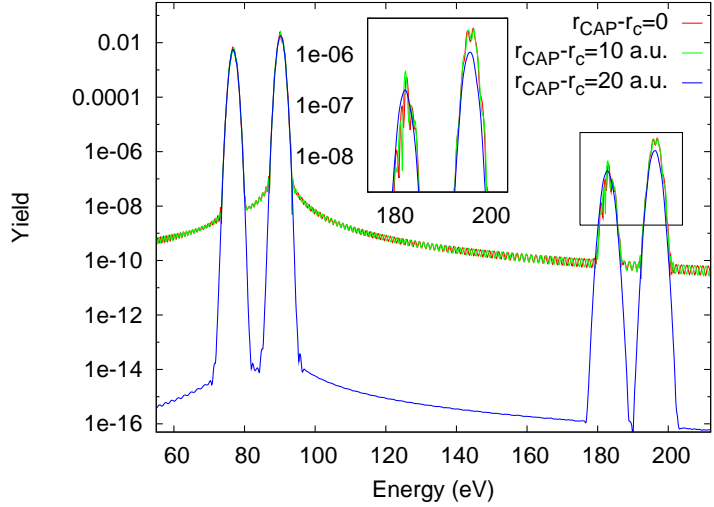}
  \caption{The argon photoelectron spectrum for the same pulse as in Figs.~\ref{fig1} to \ref{fig3} and Figs.~\ref{figtime} to \ref{fig5} is shown along the polarization direction. The numerical parameters are $r_{\rm max}=250$~a.u., $\h=1\times10^{-3}$, $r_{\rm CAP}=220$~a.u., and the propagation time is $1000$~a.u. The distance $r_{\rm max}\!-\!r_{\rm CAP}$ is varied in the range from $0$ to $20$~a.u.}
 \label{rmaxrc}
\end{figure}

The method relies on an optimized CAP for the energy range of interest. However, the CAP cannot guarantee a perfect absorption. Since the optimized CAP strength is energy dependent \cite{rismey} the tsurff spectrum can be optimized only for a limited energy range. In Fig.~\ref{fig5} the energy spectrum for $\theta=0$ is shown for different CAP strengths $\h$. It is clear that reflections from the CAP as well as from the end of the radial grid leave a trace in the spectrum. A weak CAP cannot fully absorb a fast electron before the end of the radial grid. On the other hand, a strong CAP will reflect the electron. For the kinetic energies of the electrons considered here the optimized CAP parameter lies at a value of about $10^{-3}$. Of course, the other parameter that must be optimized is the CAP radius $r_{\rm CAP}$. We find that the optimum is an absorption range of $r_{\rm max}\!-\!r_{\rm CAP}=30$~a.u. For tsurff also the distance of $r_c$ to $r_{\rm CAP}$ plays a role. In Fig.~\ref{rmaxrc} the spectrum is shown for different $r_{\rm CAP}-r_c$ values. For a distance of $r_{\rm CAP}-r_c=20$~a.u. the spectrum becomes less oscillatory and the noise level decreases significantly in comparison to shorter ranges $r_{\rm CAP}-r_c$.

\begin{figure}[htbp]
 \centering
 \includegraphics[width=10cm]{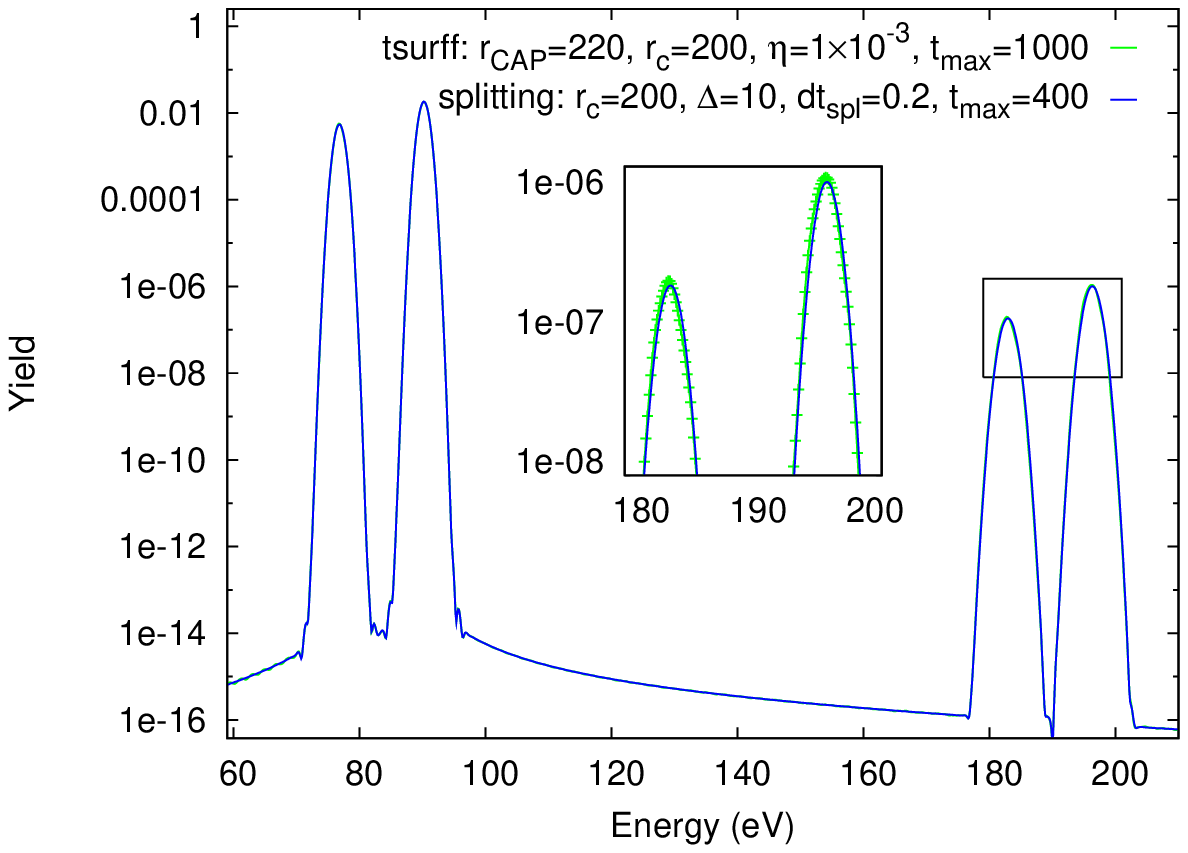}
 \caption{The argon photoelectron spectra obtained with the splitting and the tsurff methods for a pulse with $105$~eV photon energy, $9\times 10^{13}$~Wcm$^{-2}$ intensity and $1.2$~fs duration along the polarization direction are compared. The radial grid size is $r_{\rm max}=250$~a.u. for both methods.}
 \label{fig6}
\end{figure}

A direct comparison of the spectrum in the direction $\theta=0$ obtained by splitting and tsurff, respectively, is shown in Fig.~\ref{fig6}. The pulse characteristics are the same as for the Figs.~\ref{fig1} to \ref{fig3} and \ref{figtime} to \ref{rmaxrc}. The radial grid size is $r_{\rm max}=250$~a.u. The splitting parameters are $r_c=200$~a.u., $\D=10$~a.u., ${\rm d} t_{spl}=0.2$~a.u., and the propagation time is $400$~a.u. For the surface flux method a CAP strength of $10^{-3}$, a CAP radius of $r_{\rm CAP}=220$~a.u., a sphere radius of $r_c=200$~a.u. (according to the optimum found for $r_{\rm CAP}-r_c=20$~a.u.) and a propagation time of $1000$~a.u. are chosen. The spectra agree quite nicely. The one-photon peaks exhibit a nearly perfect agreement. The slight deviation in the two-photon spectrum calculated with tsurff indicates that the CAP could be reoptimized for this energy range. However, for both methods the spectrum has a very low noise level, up to ten orders of magnitude smaller than the physical signal.
Summarizing, the tsurff-method is in principle applicable with a CAP, although it requires a good quality absorption over a broad energy range. Qualitatively, the tsurff method reproduces exactly the same results as obtained with the splitting method.

\section{Conclusion}\label{conc}
We have implemented two computational methods for the calculation of photoelectron spectra within the TDCIS scheme. Both methods can produce reasonable and quantitative energy- and angle-resolved spectra within our model. Subshell ionization can be quantified. We have applied and compared these methods for the high-intensity XUV regime. Advantages of the splitting method are the good absorption characteristics through the splitting function and the short propagation time that is needed. A disadvantage is the long evaluation time of the radial integrals. The tsurff method needs a longer propagation time. However, the calculation of the photoelectron spectrum in the analysis step is much faster than with the splitting method due to the evaluation at one point. The comparison of the two methods shows that, in principle, the same spectra can be obtained after the appropriate optimization of the computational parameters.
Although our application in the present work focuses on the XUV range, it is of course possible to study also processes in the strong-field regime in the infrared range by analyzing the photoelectron spectrum. Interesting applications arise from the fact that information about the coherence and the entanglement of the ionic state and the photoelectron can be extracted from the properties of the outgoing wavepacket. 

\section{Acknowledgments}
This work has been supported by the Deutsche Forschungsgemeinschaft under Grant No. SFB 925/A5.

\bibliography{sflit.bib}

\begin{thebibliography}{73}
\expandafter\ifx\csname natexlab\endcsname\relax\def\natexlab#1{#1}\fi
\expandafter\ifx\csname bibnamefont\endcsname\relax
  \def\bibnamefont#1{#1}\fi
\expandafter\ifx\csname bibfnamefont\endcsname\relax
  \def\bibfnamefont#1{#1}\fi
\expandafter\ifx\csname citenamefont\endcsname\relax
  \def\citenamefont#1{#1}\fi
\expandafter\ifx\csname url\endcsname\relax
  \def\url#1{\texttt{#1}}\fi
\expandafter\ifx\csname urlprefix\endcsname\relax\def\urlprefix{URL }\fi
\providecommand{\bibinfo}[2]{#2}
\providecommand{\eprint}[2][]{\url{#2}}

\bibitem[{\citenamefont{Tong et~al.}(2006)\citenamefont{Tong, Hino, and
  Toshima}}]{tong}
\bibinfo{author}{\bibfnamefont{X.}~\bibnamefont{Tong}},
  \bibinfo{author}{\bibfnamefont{K.}~\bibnamefont{Hino}}, \bibnamefont{and}
  \bibinfo{author}{\bibfnamefont{N.}~\bibnamefont{Toshima}},
  \bibinfo{journal}{Phys. Rev. A} \textbf{\bibinfo{volume}{74}},
  \bibinfo{pages}{031405(R)} (\bibinfo{year}{2006}).

\bibitem[{\citenamefont{Tao and Scrinzi}(2012)}]{scrinzi}
\bibinfo{author}{\bibfnamefont{L.}~\bibnamefont{Tao}} \bibnamefont{and}
  \bibinfo{author}{\bibfnamefont{A.}~\bibnamefont{Scrinzi}},
  \bibinfo{journal}{New J. Phys.} \textbf{\bibinfo{volume}{14}},
  \bibinfo{pages}{013021} (\bibinfo{year}{2012}).

\bibitem[{\citenamefont{Corkum and Krausz}(2007)}]{corkumkrausz}
\bibinfo{author}{\bibfnamefont{P.}~\bibnamefont{Corkum}} \bibnamefont{and}
  \bibinfo{author}{\bibfnamefont{F.}~\bibnamefont{Krausz}},
  \bibinfo{journal}{Nature Physics} \textbf{\bibinfo{volume}{3}},
  \bibinfo{pages}{381} (\bibinfo{year}{2007}).

\bibitem[{\citenamefont{Chapman et~al.}(2006)\citenamefont{Chapman, Barty,
  Bogan, Boutet, Frank, Hau-Riege, Marchesini, Woods, Bajt, Benner
  et~al.}}]{chapman}
\bibinfo{author}{\bibfnamefont{H.~N.} \bibnamefont{Chapman}},
  \bibinfo{author}{\bibfnamefont{A.}~\bibnamefont{Barty}},
  \bibinfo{author}{\bibfnamefont{M.~J.} \bibnamefont{Bogan}},
  \bibinfo{author}{\bibfnamefont{S.}~\bibnamefont{Boutet}},
  \bibinfo{author}{\bibfnamefont{M.}~\bibnamefont{Frank}},
  \bibinfo{author}{\bibfnamefont{S.~P.} \bibnamefont{Hau-Riege}},
  \bibinfo{author}{\bibfnamefont{S.}~\bibnamefont{Marchesini}},
  \bibinfo{author}{\bibfnamefont{B.~W.} \bibnamefont{Woods}},
  \bibinfo{author}{\bibfnamefont{S.}~\bibnamefont{Bajt}},
  \bibinfo{author}{\bibfnamefont{W.~H.} \bibnamefont{Benner}},
  \bibnamefont{et~al.}, \bibinfo{journal}{Nature} \textbf{\bibinfo{volume}{2}},
  \bibinfo{pages}{839} (\bibinfo{year}{2006}).

\bibitem[{\citenamefont{Itatani et~al.}(2004)\citenamefont{Itatani, Levesque,
  Zeidler, Niikura, P{\'e}pin, Kieffer, Corkum, and Villeneuve}}]{levesque}
\bibinfo{author}{\bibfnamefont{J.}~\bibnamefont{Itatani}},
  \bibinfo{author}{\bibfnamefont{J.}~\bibnamefont{Levesque}},
  \bibinfo{author}{\bibfnamefont{D.}~\bibnamefont{Zeidler}},
  \bibinfo{author}{\bibfnamefont{H.}~\bibnamefont{Niikura}},
  \bibinfo{author}{\bibfnamefont{H.}~\bibnamefont{P{\'e}pin}},
  \bibinfo{author}{\bibfnamefont{J.~C.} \bibnamefont{Kieffer}},
  \bibinfo{author}{\bibfnamefont{P.~B.} \bibnamefont{Corkum}},
  \bibnamefont{and} \bibinfo{author}{\bibfnamefont{D.~M.}
  \bibnamefont{Villeneuve}}, \bibinfo{journal}{Nature}
  \textbf{\bibinfo{volume}{432}}, \bibinfo{pages}{867} (\bibinfo{year}{2004}).

\bibitem[{\citenamefont{Spanner et~al.}(2004)\citenamefont{Spanner, Smirnova,
  Corkum, and Ivanov}}]{spanner}
\bibinfo{author}{\bibfnamefont{M.}~\bibnamefont{Spanner}},
  \bibinfo{author}{\bibfnamefont{O.}~\bibnamefont{Smirnova}},
  \bibinfo{author}{\bibfnamefont{P.~B.} \bibnamefont{Corkum}},
  \bibnamefont{and} \bibinfo{author}{\bibfnamefont{M.~Y.}
  \bibnamefont{Ivanov}}, \bibinfo{journal}{J. Phys. B: At. Mol. Phys.}
  \textbf{\bibinfo{volume}{37}}, \bibinfo{pages}{243} (\bibinfo{year}{2004}).

\bibitem[{\citenamefont{Pabst}(2013)}]{stefanreview}
\bibinfo{author}{\bibfnamefont{S.}~\bibnamefont{Pabst}}, \bibinfo{journal}{The
  European Physical Journal Special Topics} \textbf{\bibinfo{volume}{221}},
  \bibinfo{pages}{1} (\bibinfo{year}{2013}).

\bibitem[{\citenamefont{Kanter et~al.}(2011)\citenamefont{Kanter, Kr{\"a}ssig,
  Li, March, Ho, Rohringer, Santra, Southworth, DiMauro, Doumy
  et~al.}}]{kanter}
\bibinfo{author}{\bibfnamefont{E.~P.} \bibnamefont{Kanter}},
  \bibinfo{author}{\bibfnamefont{B.}~\bibnamefont{Kr{\"a}ssig}},
  \bibinfo{author}{\bibfnamefont{Y.}~\bibnamefont{Li}},
  \bibinfo{author}{\bibfnamefont{A.~M.} \bibnamefont{March}},
  \bibinfo{author}{\bibfnamefont{P.}~\bibnamefont{Ho}},
  \bibinfo{author}{\bibfnamefont{N.}~\bibnamefont{Rohringer}},
  \bibinfo{author}{\bibfnamefont{R.}~\bibnamefont{Santra}},
  \bibinfo{author}{\bibfnamefont{S.~H.} \bibnamefont{Southworth}},
  \bibinfo{author}{\bibfnamefont{L.~F.} \bibnamefont{DiMauro}},
  \bibinfo{author}{\bibfnamefont{G.}~\bibnamefont{Doumy}},
  \bibnamefont{et~al.}, \bibinfo{journal}{Phys. Rev. Lett.}
  \textbf{\bibinfo{volume}{107}}, \bibinfo{pages}{233001}
  (\bibinfo{year}{2011}).

\bibitem[{\citenamefont{Demekhin and Cederbaum}(2012)}]{demekhin}
\bibinfo{author}{\bibfnamefont{P.~V.} \bibnamefont{Demekhin}} \bibnamefont{and}
  \bibinfo{author}{\bibfnamefont{L.~S.} \bibnamefont{Cederbaum}},
  \bibinfo{journal}{Phys. Rev. A} \textbf{\bibinfo{volume}{86}},
  \bibinfo{pages}{063412} (\bibinfo{year}{2012}).

\bibitem[{\citenamefont{Agostini et~al.}(1979)\citenamefont{Agostini, Fabre,
  Mainfray, and Petite}}]{ago}
\bibinfo{author}{\bibfnamefont{P.}~\bibnamefont{Agostini}},
  \bibinfo{author}{\bibfnamefont{F.}~\bibnamefont{Fabre}},
  \bibinfo{author}{\bibfnamefont{G.}~\bibnamefont{Mainfray}}, \bibnamefont{and}
  \bibinfo{author}{\bibfnamefont{G.}~\bibnamefont{Petite}},
  \bibinfo{journal}{Phys. Rev. Lett.} \textbf{\bibinfo{volume}{42}},
  \bibinfo{pages}{1127} (\bibinfo{year}{1979}).

\bibitem[{\citenamefont{Emma et~al.}(2004)\citenamefont{Emma, Bane, Cornacchia,
  Huang, Schlarb, Stupakov, and Walz}}]{emma}
\bibinfo{author}{\bibfnamefont{P.}~\bibnamefont{Emma}},
  \bibinfo{author}{\bibfnamefont{K.}~\bibnamefont{Bane}},
  \bibinfo{author}{\bibfnamefont{M.}~\bibnamefont{Cornacchia}},
  \bibinfo{author}{\bibfnamefont{Z.}~\bibnamefont{Huang}},
  \bibinfo{author}{\bibfnamefont{H.}~\bibnamefont{Schlarb}},
  \bibinfo{author}{\bibfnamefont{G.}~\bibnamefont{Stupakov}}, \bibnamefont{and}
  \bibinfo{author}{\bibfnamefont{D.}~\bibnamefont{Walz}},
  \bibinfo{journal}{Phys. Rev. Lett.} \textbf{\bibinfo{volume}{92}},
  \bibinfo{pages}{074801} (\bibinfo{year}{2004}).

\bibitem[{\citenamefont{Krausz and Ivanov}(2009)}]{kra}
\bibinfo{author}{\bibfnamefont{F.}~\bibnamefont{Krausz}} \bibnamefont{and}
  \bibinfo{author}{\bibfnamefont{M.~Y.} \bibnamefont{Ivanov}},
  \bibinfo{journal}{Rev. Mod. Phys.} \textbf{\bibinfo{volume}{81}},
  \bibinfo{pages}{163} (\bibinfo{year}{2009}).

\bibitem[{\citenamefont{Baker et~al.}(2006)\citenamefont{Baker, Robinson,
  Haworth, Teng, Smith, Chiril\v{a}, Lein, Tisch, and Marangos}}]{baker}
\bibinfo{author}{\bibfnamefont{S.}~\bibnamefont{Baker}},
  \bibinfo{author}{\bibfnamefont{J.~S.} \bibnamefont{Robinson}},
  \bibinfo{author}{\bibfnamefont{C.~A.} \bibnamefont{Haworth}},
  \bibinfo{author}{\bibfnamefont{H.}~\bibnamefont{Teng}},
  \bibinfo{author}{\bibfnamefont{R.~A.} \bibnamefont{Smith}},
  \bibinfo{author}{\bibfnamefont{C.~C.} \bibnamefont{Chiril\v{a}}},
  \bibinfo{author}{\bibfnamefont{M.}~\bibnamefont{Lein}},
  \bibinfo{author}{\bibfnamefont{J.~W.~G.} \bibnamefont{Tisch}},
  \bibnamefont{and} \bibinfo{author}{\bibfnamefont{J.~P.}
  \bibnamefont{Marangos}}, \textbf{\bibinfo{volume}{312}}, \bibinfo{pages}{424}
  (\bibinfo{year}{2006}).

\bibitem[{\citenamefont{Wabnitz et~al.}(2002)\citenamefont{Wabnitz, Bittner,
  {de Castro}, D{\"o}hrmann, G{\"u}rtler, Laarmann, Laasch, Schulz, Swiderski,
  {von Haeften} et~al.}}]{wabnitz}
\bibinfo{author}{\bibfnamefont{H.}~\bibnamefont{Wabnitz}},
  \bibinfo{author}{\bibfnamefont{L.}~\bibnamefont{Bittner}},
  \bibinfo{author}{\bibfnamefont{A.~R.~B.} \bibnamefont{{de Castro}}},
  \bibinfo{author}{\bibfnamefont{R.}~\bibnamefont{D{\"o}hrmann}},
  \bibinfo{author}{\bibfnamefont{P.}~\bibnamefont{G{\"u}rtler}},
  \bibinfo{author}{\bibfnamefont{T.}~\bibnamefont{Laarmann}},
  \bibinfo{author}{\bibfnamefont{W.}~\bibnamefont{Laasch}},
  \bibinfo{author}{\bibfnamefont{J.}~\bibnamefont{Schulz}},
  \bibinfo{author}{\bibfnamefont{A.}~\bibnamefont{Swiderski}},
  \bibinfo{author}{\bibfnamefont{K.}~\bibnamefont{{von Haeften}}},
  \bibnamefont{et~al.}, \bibinfo{journal}{Nature}
  \textbf{\bibinfo{volume}{420}}, \bibinfo{pages}{482} (\bibinfo{year}{2002}).

\bibitem[{\citenamefont{Young et~al.}(2010)\citenamefont{Young, Kanter,
  Kr{\"a}ssig, Li, March, Pratt, Santra, Southworth, Rohringer, DiMauro
  et~al.}}]{young}
\bibinfo{author}{\bibfnamefont{L.}~\bibnamefont{Young}},
  \bibinfo{author}{\bibfnamefont{E.~P.} \bibnamefont{Kanter}},
  \bibinfo{author}{\bibfnamefont{B.}~\bibnamefont{Kr{\"a}ssig}},
  \bibinfo{author}{\bibfnamefont{Y.}~\bibnamefont{Li}},
  \bibinfo{author}{\bibfnamefont{A.~M.} \bibnamefont{March}},
  \bibinfo{author}{\bibfnamefont{S.~T.} \bibnamefont{Pratt}},
  \bibinfo{author}{\bibfnamefont{R.}~\bibnamefont{Santra}},
  \bibinfo{author}{\bibfnamefont{S.~H.} \bibnamefont{Southworth}},
  \bibinfo{author}{\bibfnamefont{N.}~\bibnamefont{Rohringer}},
  \bibinfo{author}{\bibfnamefont{L.~F.} \bibnamefont{DiMauro}},
  \bibnamefont{et~al.}, \bibinfo{journal}{Nature}
  \textbf{\bibinfo{volume}{466}}, \bibinfo{pages}{56} (\bibinfo{year}{2010}).

\bibitem[{\citenamefont{Moshammer et~al.}(2007)\citenamefont{Moshammer, Jiang,
  Foucar, Rudenko, Ergler, Schr{\"o}ter, L{\"u}demann, Zrost, Fischer, Titze
  et~al.}}]{moshammer}
\bibinfo{author}{\bibfnamefont{R.}~\bibnamefont{Moshammer}},
  \bibinfo{author}{\bibfnamefont{Y.~H.} \bibnamefont{Jiang}},
  \bibinfo{author}{\bibfnamefont{L.}~\bibnamefont{Foucar}},
  \bibinfo{author}{\bibfnamefont{A.}~\bibnamefont{Rudenko}},
  \bibinfo{author}{\bibfnamefont{T.}~\bibnamefont{Ergler}},
  \bibinfo{author}{\bibfnamefont{C.~D.} \bibnamefont{Schr{\"o}ter}},
  \bibinfo{author}{\bibfnamefont{S.}~\bibnamefont{L{\"u}demann}},
  \bibinfo{author}{\bibfnamefont{K.}~\bibnamefont{Zrost}},
  \bibinfo{author}{\bibfnamefont{D.}~\bibnamefont{Fischer}},
  \bibinfo{author}{\bibfnamefont{J.}~\bibnamefont{Titze}},
  \bibnamefont{et~al.}, \bibinfo{journal}{Phys. Rev. Lett.}
  \textbf{\bibinfo{volume}{98}}, \bibinfo{pages}{203001}
  (\bibinfo{year}{2007}).

\bibitem[{\citenamefont{Sytcheva et~al.}(2012)\citenamefont{Sytcheva, Pabst,
  Son, and Santra}}]{arina}
\bibinfo{author}{\bibfnamefont{A.}~\bibnamefont{Sytcheva}},
  \bibinfo{author}{\bibfnamefont{S.}~\bibnamefont{Pabst}},
  \bibinfo{author}{\bibfnamefont{S.-K.} \bibnamefont{Son}}, \bibnamefont{and}
  \bibinfo{author}{\bibfnamefont{R.}~\bibnamefont{Santra}},
  \bibinfo{journal}{Phys. Rev. A} \textbf{\bibinfo{volume}{85}},
  \bibinfo{pages}{023414} (\bibinfo{year}{2012}).

\bibitem[{\citenamefont{Drescher et~al.}(2002)\citenamefont{Drescher,
  Hentschel, Kienberger, Uiberacker, Yakovlev, Scrinzi, Westerwalbesloh,
  Kleineberg, Heinzmann, and Krausz}}]{drescher}
\bibinfo{author}{\bibfnamefont{M.}~\bibnamefont{Drescher}},
  \bibinfo{author}{\bibfnamefont{M.}~\bibnamefont{Hentschel}},
  \bibinfo{author}{\bibfnamefont{R.}~\bibnamefont{Kienberger}},
  \bibinfo{author}{\bibfnamefont{M.}~\bibnamefont{Uiberacker}},
  \bibinfo{author}{\bibfnamefont{V.}~\bibnamefont{Yakovlev}},
  \bibinfo{author}{\bibfnamefont{A.}~\bibnamefont{Scrinzi}},
  \bibinfo{author}{\bibfnamefont{T.}~\bibnamefont{Westerwalbesloh}},
  \bibinfo{author}{\bibfnamefont{U.}~\bibnamefont{Kleineberg}},
  \bibinfo{author}{\bibfnamefont{U.}~\bibnamefont{Heinzmann}},
  \bibnamefont{and} \bibinfo{author}{\bibfnamefont{F.}~\bibnamefont{Krausz}},
  \bibinfo{journal}{Nature} \textbf{\bibinfo{volume}{419}},
  \bibinfo{pages}{803} (\bibinfo{year}{2002}).

\bibitem[{\citenamefont{Fukuzawa et~al.}(2013)\citenamefont{Fukuzawa, Son,
  Motomura, Mondal, Nagaya, Wada, Liu, Feifel, Tachibana, Ito
  et~al.}}]{fukuzawa}
\bibinfo{author}{\bibfnamefont{H.}~\bibnamefont{Fukuzawa}},
  \bibinfo{author}{\bibfnamefont{S.-K.} \bibnamefont{Son}},
  \bibinfo{author}{\bibfnamefont{K.}~\bibnamefont{Motomura}},
  \bibinfo{author}{\bibfnamefont{S.}~\bibnamefont{Mondal}},
  \bibinfo{author}{\bibfnamefont{K.}~\bibnamefont{Nagaya}},
  \bibinfo{author}{\bibfnamefont{S.}~\bibnamefont{Wada}},
  \bibinfo{author}{\bibfnamefont{X.-J.} \bibnamefont{Liu}},
  \bibinfo{author}{\bibfnamefont{R.}~\bibnamefont{Feifel}},
  \bibinfo{author}{\bibfnamefont{T.}~\bibnamefont{Tachibana}},
  \bibinfo{author}{\bibfnamefont{Y.}~\bibnamefont{Ito}}, \bibnamefont{et~al.},
  \bibinfo{journal}{Phys. Rev. Lett.} \textbf{\bibinfo{volume}{110}},
  \bibinfo{pages}{173005} (\bibinfo{year}{2013}).

\bibitem[{\citenamefont{H{\"u}fner}(1996)}]{huefner}
\bibinfo{author}{\bibfnamefont{S.}~\bibnamefont{H{\"u}fner}},
  \emph{\bibinfo{title}{Photoelectron Spectroscopy}}
  (\bibinfo{publisher}{Springer}, \bibinfo{year}{1996}).

\bibitem[{\citenamefont{Wu et~al.}(2011)\citenamefont{Wu, Hockett, and
  Stolow}}]{stolow}
\bibinfo{author}{\bibfnamefont{G.}~\bibnamefont{Wu}},
  \bibinfo{author}{\bibfnamefont{P.}~\bibnamefont{Hockett}}, \bibnamefont{and}
  \bibinfo{author}{\bibfnamefont{A.}~\bibnamefont{Stolow}},
  \bibinfo{journal}{Phys. Chem. Chem. Phys.} \textbf{\bibinfo{volume}{13}},
  \bibinfo{pages}{18447} (\bibinfo{year}{2011}).

\bibitem[{\citenamefont{Fabre et~al.}(1981)\citenamefont{Fabre, Agostini,
  Petite, and Clement}}]{fabre}
\bibinfo{author}{\bibfnamefont{F.}~\bibnamefont{Fabre}},
  \bibinfo{author}{\bibfnamefont{P.}~\bibnamefont{Agostini}},
  \bibinfo{author}{\bibfnamefont{G.}~\bibnamefont{Petite}}, \bibnamefont{and}
  \bibinfo{author}{\bibfnamefont{M.}~\bibnamefont{Clement}},
  \bibinfo{journal}{J. Phys. B: At. Mol. Phys.} \textbf{\bibinfo{volume}{14}},
  \bibinfo{pages}{L677} (\bibinfo{year}{1981}).

\bibitem[{\citenamefont{Schmidt}(1986)}]{schmidt}
\bibinfo{author}{\bibfnamefont{V.}~\bibnamefont{Schmidt}},
  \bibinfo{journal}{Zeitschrift f{\"u}r Physik D Atoms, Molecules and Clusters}
  \textbf{\bibinfo{volume}{2}}, \bibinfo{pages}{275} (\bibinfo{year}{1986}).

\bibitem[{\citenamefont{Becker and Shirley}(1990)}]{uwebecker}
\bibinfo{author}{\bibfnamefont{U.}~\bibnamefont{Becker}} \bibnamefont{and}
  \bibinfo{author}{\bibfnamefont{D.~A.} \bibnamefont{Shirley}},
  \bibinfo{journal}{Phys. Scr. T} \textbf{\bibinfo{volume}{31}},
  \bibinfo{pages}{56} (\bibinfo{year}{1990}).

\bibitem[{\citenamefont{Meyer et~al.}(2010)\citenamefont{Meyer, Cubaynes,
  Richardson, Costello, Radcliffe, Li, D{\"u}sterer, Fritzsche, Mihelic,
  Papamihail et~al.}}]{meyerpapa}
\bibinfo{author}{\bibfnamefont{M.}~\bibnamefont{Meyer}},
  \bibinfo{author}{\bibfnamefont{D.}~\bibnamefont{Cubaynes}},
  \bibinfo{author}{\bibfnamefont{V.}~\bibnamefont{Richardson}},
  \bibinfo{author}{\bibfnamefont{J.~T.} \bibnamefont{Costello}},
  \bibinfo{author}{\bibfnamefont{P.}~\bibnamefont{Radcliffe}},
  \bibinfo{author}{\bibfnamefont{W.~B.} \bibnamefont{Li}},
  \bibinfo{author}{\bibfnamefont{S.}~\bibnamefont{D{\"u}sterer}},
  \bibinfo{author}{\bibfnamefont{S.}~\bibnamefont{Fritzsche}},
  \bibinfo{author}{\bibfnamefont{A.}~\bibnamefont{Mihelic}},
  \bibinfo{author}{\bibfnamefont{K.~G.} \bibnamefont{Papamihail}},
  \bibnamefont{et~al.}, \bibinfo{journal}{Phys. Rev. Lett.}
  \textbf{\bibinfo{volume}{104}}, \bibinfo{pages}{213001}
  (\bibinfo{year}{2010}).

\bibitem[{\citenamefont{Blaga et~al.}(2008)\citenamefont{Blaga, Catoire,
  Colosimo, Paulus, Muller, Agostini, and DiMauro}}]{blaga}
\bibinfo{author}{\bibfnamefont{C.~I.} \bibnamefont{Blaga}},
  \bibinfo{author}{\bibfnamefont{F.}~\bibnamefont{Catoire}},
  \bibinfo{author}{\bibfnamefont{P.}~\bibnamefont{Colosimo}},
  \bibinfo{author}{\bibfnamefont{G.~G.} \bibnamefont{Paulus}},
  \bibinfo{author}{\bibfnamefont{H.~G.} \bibnamefont{Muller}},
  \bibinfo{author}{\bibfnamefont{P.}~\bibnamefont{Agostini}}, \bibnamefont{and}
  \bibinfo{author}{\bibfnamefont{L.~F.} \bibnamefont{DiMauro}},
  \bibinfo{journal}{Nature Physics} \textbf{\bibinfo{volume}{5}},
  \bibinfo{pages}{335} (\bibinfo{year}{2008}).

\bibitem[{\citenamefont{Quan et~al.}(2009)\citenamefont{Quan, Lin, Wu, Kang,
  Liu, Liu, Chen, Liu, He, Chen et~al.}}]{quan}
\bibinfo{author}{\bibfnamefont{W.}~\bibnamefont{Quan}},
  \bibinfo{author}{\bibfnamefont{Z.}~\bibnamefont{Lin}},
  \bibinfo{author}{\bibfnamefont{M.}~\bibnamefont{Wu}},
  \bibinfo{author}{\bibfnamefont{H.}~\bibnamefont{Kang}},
  \bibinfo{author}{\bibfnamefont{H.}~\bibnamefont{Liu}},
  \bibinfo{author}{\bibfnamefont{X.}~\bibnamefont{Liu}},
  \bibinfo{author}{\bibfnamefont{J.}~\bibnamefont{Chen}},
  \bibinfo{author}{\bibfnamefont{J.}~\bibnamefont{Liu}},
  \bibinfo{author}{\bibfnamefont{X.~T.} \bibnamefont{He}},
  \bibinfo{author}{\bibfnamefont{S.~G.} \bibnamefont{Chen}},
  \bibnamefont{et~al.}, \bibinfo{journal}{Phys. Rev. Lett.}
  \textbf{\bibinfo{volume}{103}}, \bibinfo{pages}{093001}
  (\bibinfo{year}{2009}).

\bibitem[{\citenamefont{Bethe and Salpeter}(1977)}]{bethe}
\bibinfo{author}{\bibfnamefont{H.}~\bibnamefont{Bethe}} \bibnamefont{and}
  \bibinfo{author}{\bibfnamefont{E.}~\bibnamefont{Salpeter}},
  \emph{\bibinfo{title}{Quantum Mechanics of One-- and Two--Electron Atoms}}
  (\bibinfo{publisher}{Springer}, \bibinfo{year}{1977}).

\bibitem[{\citenamefont{Starace}(1982)}]{atomicphoto}
\bibinfo{author}{\bibfnamefont{A.}~\bibnamefont{Starace}},
  \emph{\bibinfo{title}{Theory of Atomic Photoionization}}
  (\bibinfo{publisher}{Springer}, \bibinfo{year}{1982}),
  vol.~\bibinfo{volume}{31}, pp. \bibinfo{pages}{1--121}.

\bibitem[{\citenamefont{Krause et~al.}(1992)\citenamefont{Krause, Schafer, and
  Kulander}}]{krause}
\bibinfo{author}{\bibfnamefont{J.~L.} \bibnamefont{Krause}},
  \bibinfo{author}{\bibfnamefont{K.~J.} \bibnamefont{Schafer}},
  \bibnamefont{and} \bibinfo{author}{\bibfnamefont{K.~C.}
  \bibnamefont{Kulander}}, \bibinfo{journal}{Phys. Rev. A}
  \textbf{\bibinfo{volume}{45}}, \bibinfo{pages}{4998} (\bibinfo{year}{1992}).

\bibitem[{\citenamefont{M. et~al.}(2000)\citenamefont{M., N.J., and
  C.J.}}]{joachain}
\bibinfo{author}{\bibfnamefont{D.}~\bibnamefont{M.}},
  \bibinfo{author}{\bibfnamefont{K.}~\bibnamefont{N.J.}}, \bibnamefont{and}
  \bibinfo{author}{\bibfnamefont{J.}~\bibnamefont{C.J.}},
  \bibinfo{journal}{Adv. At. Mol. Opt. Phys.} \textbf{\bibinfo{volume}{42}},
  \bibinfo{pages}{225} (\bibinfo{year}{2000}).

\bibitem[{\citenamefont{Chelkowski et~al.}(2004)\citenamefont{Chelkowski,
  Bandrauk, and Apolonski}}]{chelk}
\bibinfo{author}{\bibfnamefont{S.}~\bibnamefont{Chelkowski}},
  \bibinfo{author}{\bibfnamefont{A.~D.} \bibnamefont{Bandrauk}},
  \bibnamefont{and}
  \bibinfo{author}{\bibfnamefont{A.}~\bibnamefont{Apolonski}},
  \bibinfo{journal}{Phys. Rev. A} \textbf{\bibinfo{volume}{70}},
  \bibinfo{pages}{013815} (\bibinfo{year}{2004}).

\bibitem[{\citenamefont{Starace}(1970)}]{starace}
\bibinfo{author}{\bibfnamefont{A.~F.} \bibnamefont{Starace}},
  \bibinfo{journal}{Phys. Rev. A} \textbf{\bibinfo{volume}{2}},
  \bibinfo{pages}{118} (\bibinfo{year}{1970}).

\bibitem[{\citenamefont{{van der Hart} and Greene}(1998)}]{hugovan}
\bibinfo{author}{\bibfnamefont{H.~W.} \bibnamefont{{van der Hart}}}
  \bibnamefont{and} \bibinfo{author}{\bibfnamefont{C.~H.}
  \bibnamefont{Greene}}, \bibinfo{journal}{Phys. Rev. A}
  \textbf{\bibinfo{volume}{58}}, \bibinfo{pages}{2097} (\bibinfo{year}{1998}).

\bibitem[{\citenamefont{Shavitt}(1977)}]{shavitt}
\bibinfo{author}{\bibfnamefont{I.}~\bibnamefont{Shavitt}},
  \emph{\bibinfo{title}{The method of configuration interaction}}
  (\bibinfo{publisher}{Plenum Press}, \bibinfo{year}{1977}).

\bibitem[{\citenamefont{K{\"u}mmel}(2003)}]{kuemmel}
\bibinfo{author}{\bibfnamefont{H.}~\bibnamefont{K{\"u}mmel}},
  \bibinfo{journal}{Int. J. Mod. Phys. B} \textbf{\bibinfo{volume}{17}},
  \bibinfo{pages}{5311} (\bibinfo{year}{2003}).

\bibitem[{\citenamefont{Amusia et~al.}(2012)\citenamefont{Amusia, Chernysheva,
  and Yarzhemsky}}]{amusia}
\bibinfo{author}{\bibfnamefont{M.}~\bibnamefont{Amusia}},
  \bibinfo{author}{\bibfnamefont{L.}~\bibnamefont{Chernysheva}},
  \bibnamefont{and}
  \bibinfo{author}{\bibfnamefont{V.}~\bibnamefont{Yarzhemsky}},
  \emph{\bibinfo{title}{Handbook of Theoretical Atomic Physics}}
  (\bibinfo{publisher}{Springer}, \bibinfo{year}{2012}).

\bibitem[{\citenamefont{Dahlstr{\"o}m et~al.}(2012)\citenamefont{Dahlstr{\"o}m,
  L'Huillier, and Maquet}}]{dahl}
\bibinfo{author}{\bibfnamefont{J.~M.} \bibnamefont{Dahlstr{\"o}m}},
  \bibinfo{author}{\bibfnamefont{A.}~\bibnamefont{L'Huillier}},
  \bibnamefont{and} \bibinfo{author}{\bibfnamefont{A.}~\bibnamefont{Maquet}},
  \bibinfo{journal}{J. Phys. B: At. Mol. Phys.} \textbf{\bibinfo{volume}{45}},
  \bibinfo{pages}{183001} (\bibinfo{year}{2012}).

\bibitem[{\citenamefont{Kheifets and Ivanov}(2010)}]{delay}
\bibinfo{author}{\bibfnamefont{A.~S.} \bibnamefont{Kheifets}} \bibnamefont{and}
  \bibinfo{author}{\bibfnamefont{I.~A.} \bibnamefont{Ivanov}},
  \bibinfo{journal}{Phys. Rev. Lett.} \textbf{\bibinfo{volume}{105}},
  \bibinfo{pages}{233002} (\bibinfo{year}{2010}).

\bibitem[{\citenamefont{Burke and Taylor}(1975)}]{burke}
\bibinfo{author}{\bibfnamefont{P.~G.} \bibnamefont{Burke}} \bibnamefont{and}
  \bibinfo{author}{\bibfnamefont{K.~T.} \bibnamefont{Taylor}},
  \bibinfo{journal}{J. Phys. B: At. Mol. Phys.} \textbf{\bibinfo{volume}{8}},
  \bibinfo{pages}{2620} (\bibinfo{year}{1975}).

\bibitem[{\citenamefont{Burke and Tennyson}(2005)}]{burkelate}
\bibinfo{author}{\bibfnamefont{P.~G.} \bibnamefont{Burke}} \bibnamefont{and}
  \bibinfo{author}{\bibfnamefont{J.}~\bibnamefont{Tennyson}},
  \bibinfo{journal}{Mol. Phys.} \textbf{\bibinfo{volume}{103}},
  \bibinfo{pages}{2537} (\bibinfo{year}{2005}).

\bibitem[{\citenamefont{Taylor}(1977)}]{taylor}
\bibinfo{author}{\bibfnamefont{K.~T.} \bibnamefont{Taylor}},
  \bibinfo{journal}{J. Phys. B: At. Mol. Phys.} \textbf{\bibinfo{volume}{10}},
  \bibinfo{pages}{L699} (\bibinfo{year}{1977}).

\bibitem[{\citenamefont{Schafer and Kulander}(1990)}]{schaferkul}
\bibinfo{author}{\bibfnamefont{K.~J.} \bibnamefont{Schafer}} \bibnamefont{and}
  \bibinfo{author}{\bibfnamefont{K.~C.} \bibnamefont{Kulander}},
  \bibinfo{journal}{Phys. Rev. A} \textbf{\bibinfo{volume}{42}},
  \bibinfo{pages}{5794} (\bibinfo{year}{1990}).

\bibitem[{\citenamefont{Popruzhenko and Bauer}(2008)}]{bauer}
\bibinfo{author}{\bibfnamefont{S.}~\bibnamefont{Popruzhenko}} \bibnamefont{and}
  \bibinfo{author}{\bibfnamefont{D.}~\bibnamefont{Bauer}},
  \bibinfo{journal}{Journal of Modern Optics} \textbf{\bibinfo{volume}{55}},
  \bibinfo{pages}{2573} (\bibinfo{year}{2008}).

\bibitem[{\citenamefont{Popruzhenko et~al.}(2008)\citenamefont{Popruzhenko,
  Paulus, and Bauer}}]{popruzhenko}
\bibinfo{author}{\bibfnamefont{S.}~\bibnamefont{Popruzhenko}},
  \bibinfo{author}{\bibfnamefont{G.~G.} \bibnamefont{Paulus}},
  \bibnamefont{and} \bibinfo{author}{\bibfnamefont{D.}~\bibnamefont{Bauer}},
  \bibinfo{journal}{Physical Review A} \textbf{\bibinfo{volume}{77}},
  \bibinfo{pages}{053409} (\bibinfo{year}{2008}).

\bibitem[{\citenamefont{Burke and Burke}(1997)}]{burkeburke}
\bibinfo{author}{\bibfnamefont{P.}~\bibnamefont{Burke}} \bibnamefont{and}
  \bibinfo{author}{\bibfnamefont{V.}~\bibnamefont{Burke}}, \bibinfo{journal}{J.
  Phys. B: At. Mol. Opt. Phys.} \textbf{\bibinfo{volume}{30}},
  \bibinfo{pages}{L383} (\bibinfo{year}{1997}).

\bibitem[{\citenamefont{{van der Hart} et~al.}(2008)\citenamefont{{van der
  Hart}, Lysaght, and Burke}}]{vander}
\bibinfo{author}{\bibfnamefont{H.~W.} \bibnamefont{{van der Hart}}},
  \bibinfo{author}{\bibfnamefont{M.~A.} \bibnamefont{Lysaght}},
  \bibnamefont{and} \bibinfo{author}{\bibfnamefont{P.~G.} \bibnamefont{Burke}},
  \bibinfo{journal}{Phys. Rev. A} \textbf{\bibinfo{volume}{77}},
  \bibinfo{pages}{065401} (\bibinfo{year}{2008}).

\bibitem[{\citenamefont{Lysaght et~al.}(2009)\citenamefont{Lysaght, {van der
  Hart}, and Burke}}]{lysa}
\bibinfo{author}{\bibfnamefont{M.~A.} \bibnamefont{Lysaght}},
  \bibinfo{author}{\bibfnamefont{H.~W.} \bibnamefont{{van der Hart}}},
  \bibnamefont{and} \bibinfo{author}{\bibfnamefont{P.~G.} \bibnamefont{Burke}},
  \bibinfo{journal}{Phys. Rev. A} \textbf{\bibinfo{volume}{79}},
  \bibinfo{pages}{053411} (\bibinfo{year}{2009}).

\bibitem[{\citenamefont{Moore et~al.}(2011)\citenamefont{Moore, Lysaght,
  Nikolopoulos, Parker, {van der Hart}, and Taylor}}]{timedelay}
\bibinfo{author}{\bibfnamefont{L.}~\bibnamefont{Moore}},
  \bibinfo{author}{\bibfnamefont{M.}~\bibnamefont{Lysaght}},
  \bibinfo{author}{\bibfnamefont{L.}~\bibnamefont{Nikolopoulos}},
  \bibinfo{author}{\bibfnamefont{J.}~\bibnamefont{Parker}},
  \bibinfo{author}{\bibfnamefont{H.}~\bibnamefont{{van der Hart}}},
  \bibnamefont{and} \bibinfo{author}{\bibfnamefont{K.}~\bibnamefont{Taylor}},
  \bibinfo{journal}{J. Mod. Opt.} \textbf{\bibinfo{volume}{58}},
  \bibinfo{pages}{1132} (\bibinfo{year}{2011}).

\bibitem[{\citenamefont{Torlina et~al.}(2012)\citenamefont{Torlina, Ivanov,
  Walters, and Smirnova}}]{torlina2}
\bibinfo{author}{\bibfnamefont{L.}~\bibnamefont{Torlina}},
  \bibinfo{author}{\bibfnamefont{M.}~\bibnamefont{Ivanov}},
  \bibinfo{author}{\bibfnamefont{Z.~B.} \bibnamefont{Walters}},
  \bibnamefont{and} \bibinfo{author}{\bibfnamefont{O.}~\bibnamefont{Smirnova}},
  \bibinfo{journal}{Phys. Rev. A} \textbf{\bibinfo{volume}{86}},
  \bibinfo{pages}{043409} (\bibinfo{year}{2012}).

\bibitem[{\citenamefont{{Ngoko Djiokap} and Starace}(2011)}]{djiokap}
\bibinfo{author}{\bibfnamefont{J.~M.} \bibnamefont{{Ngoko Djiokap}}}
  \bibnamefont{and} \bibinfo{author}{\bibfnamefont{A.~F.}
  \bibnamefont{Starace}}, \bibinfo{journal}{Phys. Rev. A}
  \textbf{\bibinfo{volume}{84}}, \bibinfo{pages}{013404}
  (\bibinfo{year}{2011}).

\bibitem[{\citenamefont{Tarana and Greene}(2012)}]{tarana}
\bibinfo{author}{\bibfnamefont{M.}~\bibnamefont{Tarana}} \bibnamefont{and}
  \bibinfo{author}{\bibfnamefont{C.~H.} \bibnamefont{Greene}},
  \bibinfo{journal}{Phys. Rev. A} \textbf{\bibinfo{volume}{85}},
  \bibinfo{pages}{013411} (\bibinfo{year}{2012}).

\bibitem[{\citenamefont{Argenti et~al.}(2013)\citenamefont{Argenti, Pazourek,
  Feist, Nagele, Liertzer, Persson, Burgd{\"o}rfer, and
  Lindroth}}]{burgdoerfer}
\bibinfo{author}{\bibfnamefont{L.}~\bibnamefont{Argenti}},
  \bibinfo{author}{\bibfnamefont{R.}~\bibnamefont{Pazourek}},
  \bibinfo{author}{\bibfnamefont{J.}~\bibnamefont{Feist}},
  \bibinfo{author}{\bibfnamefont{S.}~\bibnamefont{Nagele}},
  \bibinfo{author}{\bibfnamefont{M.}~\bibnamefont{Liertzer}},
  \bibinfo{author}{\bibfnamefont{E.}~\bibnamefont{Persson}},
  \bibinfo{author}{\bibfnamefont{J.}~\bibnamefont{Burgd{\"o}rfer}},
  \bibnamefont{and} \bibinfo{author}{\bibfnamefont{E.}~\bibnamefont{Lindroth}},
  \bibinfo{journal}{Phys. Rev. A} \textbf{\bibinfo{volume}{87}},
  \bibinfo{pages}{053405} (\bibinfo{year}{2013}).

\bibitem[{\citenamefont{Hochstuhl and Bonitz}(2012)}]{hochstuhl}
\bibinfo{author}{\bibfnamefont{D.}~\bibnamefont{Hochstuhl}} \bibnamefont{and}
  \bibinfo{author}{\bibfnamefont{M.}~\bibnamefont{Bonitz}},
  \bibinfo{journal}{Phys. Rev. A} \textbf{\bibinfo{volume}{86}},
  \bibinfo{pages}{053424} (\bibinfo{year}{2012}).

\bibitem[{\citenamefont{Miyagi and Madsen}(2013)}]{miyagi}
\bibinfo{author}{\bibfnamefont{H.}~\bibnamefont{Miyagi}} \bibnamefont{and}
  \bibinfo{author}{\bibfnamefont{L.~B.} \bibnamefont{Madsen}},
  \bibinfo{journal}{Phys. Rev. A} \textbf{\bibinfo{volume}{87}},
  \bibinfo{pages}{062511} (\bibinfo{year}{2013}).

\bibitem[{\citenamefont{Feuerstein and Thumm}(2003)}]{feuer}
\bibinfo{author}{\bibfnamefont{B.}~\bibnamefont{Feuerstein}} \bibnamefont{and}
  \bibinfo{author}{\bibfnamefont{U.}~\bibnamefont{Thumm}}, \bibinfo{journal}{J.
  Phys. B: At. Mol. Opt. Phys.} \textbf{\bibinfo{volume}{36}},
  \bibinfo{pages}{707} (\bibinfo{year}{2003}).

\bibitem[{\citenamefont{Keller}(1995)}]{kellera}
\bibinfo{author}{\bibfnamefont{A.}~\bibnamefont{Keller}},
  \bibinfo{journal}{Phys. Rev. A} \textbf{\bibinfo{volume}{52}},
  \bibinfo{pages}{1450} (\bibinfo{year}{1995}).

\bibitem[{\citenamefont{Henkel et~al.}(2011)\citenamefont{Henkel, Lein, and
  Engel}}]{henkel}
\bibinfo{author}{\bibfnamefont{J.}~\bibnamefont{Henkel}},
  \bibinfo{author}{\bibfnamefont{M.}~\bibnamefont{Lein}}, \bibnamefont{and}
  \bibinfo{author}{\bibfnamefont{V.}~\bibnamefont{Engel}},
  \bibinfo{journal}{Phys. Rev. A} \textbf{\bibinfo{volume}{83}},
  \bibinfo{pages}{051401} (\bibinfo{year}{2011}).

\bibitem[{\citenamefont{Yue and Madsen}(2013)}]{yue}
\bibinfo{author}{\bibfnamefont{L.}~\bibnamefont{Yue}} \bibnamefont{and}
  \bibinfo{author}{\bibfnamefont{L.~B.} \bibnamefont{Madsen}},
  \bibinfo{journal}{Phys. Rev. A} \textbf{\bibinfo{volume}{88}},
  \bibinfo{pages}{063420} (\bibinfo{year}{2013}).

\bibitem[{\citenamefont{Greenman et~al.}(2010)\citenamefont{Greenman, Ho,
  Pabst, Kamarchik, Mazziotti, and Santra}}]{pab}
\bibinfo{author}{\bibfnamefont{L.}~\bibnamefont{Greenman}},
  \bibinfo{author}{\bibfnamefont{P.}~\bibnamefont{Ho}},
  \bibinfo{author}{\bibfnamefont{S.}~\bibnamefont{Pabst}},
  \bibinfo{author}{\bibfnamefont{E.}~\bibnamefont{Kamarchik}},
  \bibinfo{author}{\bibfnamefont{D.}~\bibnamefont{Mazziotti}},
  \bibnamefont{and} \bibinfo{author}{\bibfnamefont{R.}~\bibnamefont{Santra}},
  \bibinfo{journal}{Phys. Rev. A} \textbf{\bibinfo{volume}{82}}
  (\bibinfo{year}{2010}).

\bibitem[{\citenamefont{Pabst et~al.}(Rev 892, 2013)\citenamefont{Pabst,
  Greenman, and Santra}}]{progxcid}
\bibinfo{author}{\bibfnamefont{S.}~\bibnamefont{Pabst}},
  \bibinfo{author}{\bibfnamefont{L.}~\bibnamefont{Greenman}}, \bibnamefont{and}
  \bibinfo{author}{\bibfnamefont{R.}~\bibnamefont{Santra}},
  \emph{\bibinfo{title}{\textsc{XCID} program package for multichannel
  ionization dynamics}} (\bibinfo{year}{Rev 892, 2013}).

\bibitem[{\citenamefont{Pabst et~al.}(2011)\citenamefont{Pabst, Greenman, Ho,
  Mazziotti, and Santra}}]{decoherence}
\bibinfo{author}{\bibfnamefont{S.}~\bibnamefont{Pabst}},
  \bibinfo{author}{\bibfnamefont{L.}~\bibnamefont{Greenman}},
  \bibinfo{author}{\bibfnamefont{P.~J.} \bibnamefont{Ho}},
  \bibinfo{author}{\bibfnamefont{D.~A.} \bibnamefont{Mazziotti}},
  \bibnamefont{and} \bibinfo{author}{\bibfnamefont{R.}~\bibnamefont{Santra}},
  \bibinfo{journal}{Phys. Rev. Lett.} \textbf{\bibinfo{volume}{106}},
  \bibinfo{pages}{053003} (\bibinfo{year}{2011}).

\bibitem[{\citenamefont{Pabst and Santra}(2013)}]{giant}
\bibinfo{author}{\bibfnamefont{S.}~\bibnamefont{Pabst}} \bibnamefont{and}
  \bibinfo{author}{\bibfnamefont{R.}~\bibnamefont{Santra}},
  \bibinfo{journal}{Phys. Rev. Lett.} \textbf{\bibinfo{volume}{111}},
  \bibinfo{pages}{233005} (\bibinfo{year}{2013}).

\bibitem[{\citenamefont{Pabst et~al.}(2012{\natexlab{a}})\citenamefont{Pabst,
  Greenman, Mazziotti, and Santra}}]{coopermin}
\bibinfo{author}{\bibfnamefont{S.}~\bibnamefont{Pabst}},
  \bibinfo{author}{\bibfnamefont{L.}~\bibnamefont{Greenman}},
  \bibinfo{author}{\bibfnamefont{D.~A.} \bibnamefont{Mazziotti}},
  \bibnamefont{and} \bibinfo{author}{\bibfnamefont{R.}~\bibnamefont{Santra}},
  \bibinfo{journal}{Phys. Rev. A} \textbf{\bibinfo{volume}{85}},
  \bibinfo{pages}{023411} (\bibinfo{year}{2012}{\natexlab{a}}).

\bibitem[{\citenamefont{Pabst et~al.}(2012{\natexlab{b}})\citenamefont{Pabst,
  Sytcheva, Moulet, Wirth, Goulielmakis, and Santra}}]{transient}
\bibinfo{author}{\bibfnamefont{S.}~\bibnamefont{Pabst}},
  \bibinfo{author}{\bibfnamefont{A.}~\bibnamefont{Sytcheva}},
  \bibinfo{author}{\bibfnamefont{A.}~\bibnamefont{Moulet}},
  \bibinfo{author}{\bibfnamefont{A.}~\bibnamefont{Wirth}},
  \bibinfo{author}{\bibfnamefont{E.}~\bibnamefont{Goulielmakis}},
  \bibnamefont{and} \bibinfo{author}{\bibfnamefont{R.}~\bibnamefont{Santra}},
  \bibinfo{journal}{Phys. Rev. A} \textbf{\bibinfo{volume}{86}},
  \bibinfo{pages}{063411} (\bibinfo{year}{2012}{\natexlab{b}}).

\bibitem[{\citenamefont{Karamatskou et~al.}(2013)\citenamefont{Karamatskou,
  Pabst, and Santra}}]{antonia1}
\bibinfo{author}{\bibfnamefont{A.}~\bibnamefont{Karamatskou}},
  \bibinfo{author}{\bibfnamefont{S.}~\bibnamefont{Pabst}}, \bibnamefont{and}
  \bibinfo{author}{\bibfnamefont{R.}~\bibnamefont{Santra}},
  \bibinfo{journal}{Phys. Rev. A} \textbf{\bibinfo{volume}{87}},
  \bibinfo{pages}{043422} (\bibinfo{year}{2013}).

\bibitem[{\citenamefont{Rohringer et~al.}(2006)\citenamefont{Rohringer, Gordon,
  and Santra}}]{roh}
\bibinfo{author}{\bibfnamefont{N.}~\bibnamefont{Rohringer}},
  \bibinfo{author}{\bibfnamefont{A.}~\bibnamefont{Gordon}}, \bibnamefont{and}
  \bibinfo{author}{\bibfnamefont{R.}~\bibnamefont{Santra}},
  \bibinfo{journal}{Phys. Rev. A} \textbf{\bibinfo{volume}{74}}
  (\bibinfo{year}{2006}).

\bibitem[{\citenamefont{Scrinzi}(2010)}]{perfectabs}
\bibinfo{author}{\bibfnamefont{A.}~\bibnamefont{Scrinzi}},
  \bibinfo{journal}{Phys. Rev. A} \textbf{\bibinfo{volume}{81}},
  \bibinfo{pages}{053845} (\bibinfo{year}{2010}).

\bibitem[{\citenamefont{Scrinzi}(2012)}]{tsurff}
\bibinfo{author}{\bibfnamefont{A.}~\bibnamefont{Scrinzi}},
  \bibinfo{journal}{New J. Phys.} \textbf{\bibinfo{volume}{14}},
  \bibinfo{pages}{085008} (\bibinfo{year}{2012}).

\bibitem[{\citenamefont{Riss and Meyer}(1993)}]{rismey}
\bibinfo{author}{\bibfnamefont{U.~V.} \bibnamefont{Riss}} \bibnamefont{and}
  \bibinfo{author}{\bibfnamefont{H.-D.} \bibnamefont{Meyer}},
  \bibinfo{journal}{J. Phys. B: At. Mol. Opt. Phys.}
  \textbf{\bibinfo{volume}{26}}, \bibinfo{pages}{4503} (\bibinfo{year}{1993}).

\bibitem[{\citenamefont{Santra and Cederbaum}(2002)}]{cedersantra}
\bibinfo{author}{\bibfnamefont{R.}~\bibnamefont{Santra}} \bibnamefont{and}
  \bibinfo{author}{\bibfnamefont{L.~S.} \bibnamefont{Cederbaum}},
  \bibinfo{journal}{Phys. Rep.} \textbf{\bibinfo{volume}{368}},
  \bibinfo{pages}{1} (\bibinfo{year}{2002}).

\bibitem[{\citenamefont{Muga et~al.}(2004)\citenamefont{Muga, Palao, Navarro,
  and Egusquiza}}]{muga}
\bibinfo{author}{\bibfnamefont{J.}~\bibnamefont{Muga}},
  \bibinfo{author}{\bibfnamefont{J.}~\bibnamefont{Palao}},
  \bibinfo{author}{\bibfnamefont{B.}~\bibnamefont{Navarro}}, \bibnamefont{and}
  \bibinfo{author}{\bibfnamefont{I.}~\bibnamefont{Egusquiza}},
  \bibinfo{journal}{Phys. Rep.} \textbf{\bibinfo{volume}{395}},
  \bibinfo{pages}{357} (\bibinfo{year}{2004}).

\bibitem[{mey()}]{meyerexp}
\emph{\bibinfo{title}{Experiment carried out at {FLASH} by the group of
  {M}ichael {M}eyer ({E}uropean {XFEL}). {P}rivate communications, results to
  be published.}}

\end{thebibliography}

\end{document}